\documentclass[aps,prxquantum,reprint,showpacs,floatfix,longbibliography]{revtex4-1} 
\usepackage{graphicx} 
\usepackage{dcolumn} 
\usepackage{amssymb}
\usepackage{amsmath}
\usepackage{physics}
\usepackage{natbib}
\usepackage{xcolor}
\usepackage{soul}
\usepackage{siunitx}
\usepackage{float}
\usepackage{esdiff}
\usepackage{bbold}
\usepackage{mathtools}

\usepackage[export]{adjustbox}

\colorlet{RED}{red}

\begin{document}

\title{Emission of photon multiplets by a dc-biased superconducting circuit}

\author{G. C. \surname{M\'enard}$^{1,\dagger}$}
\email{email: gerbold.menard@gmail.com}

\author{A. \surname{Peugeot}$^{1}$}
\thanks{These two authors contributed equally.}

\author{C. \surname{Padurariu}$^{2}$}

\author{C. \surname{Rolland}$^{1}$}

\author{B. \surname{Kubala}$^{2, 3}$}

\author{Y. \surname{Mukharsky}$^{1}$}

\author{Z. \surname{Iftikhar}$^{1}$}

\author{C. \surname{Altimiras}$^{1}$}

\author{P. \surname{Roche}$^{1}$}

\author{H. \surname{le Sueur}$^{1}$}

\author{P. \surname{Joyez}$^{1}$}

\author{D. \surname{Vion}$^{1}$}

\author{D. \surname{Esteve}$^{1}$}

\author{J. \surname{Ankerhold}$^{2}$}
\email{email: joachim.ankerhold@uni-ulm.de}

\author{F. \surname{Portier}$^{1}$}
\altaffiliation[]{Deceased.}

\affiliation{$^{1}$ DSM/IRAMIS/SPEC, CNRS UMR 3680, CEA, Universit\'e Paris-Saclay, 91190 Gif sur Yvette, France}
\affiliation{$^{2}$ Institute for Complex Quantum Systems and IQST, University of Ulm, 89069 Ulm, Germany}
\affiliation{$^{3}$ Institut of Quantum Techhnologies, German Aerospace Center (DLR), 89069 Ulm, Germany}

\begin{abstract}
We observe the emission of bunches of $k \geqslant 1$ photons by a circuit made of a microwave resonator in series with a voltage-biased tunable  Josephson junction. The bunches are emitted at specific values $V_k$ of the bias voltage, for which each Cooper pair tunneling across the junction creates exactly k photons in the resonator. The latter is a micro-fabricated spiral coil which resonates and leaks photons at 4.4~GHz in a measurement line. Its characteristic impedance of 1.97~k$\Omega$ is high enough to reach a strong junction-resonator coupling and a bright emission of the k-photon bunches. We show that a RWA treatment of the system accounts quantitatively for the observed radiation intensity, from $k=1$ to $6$, and over three orders of magnitude when varying the Josephson energy $E_J$. We also measure the second order correlation function of the radiated microwave to determine its Fano factor $F_k$, which in the low $E_J$ limit, confirms with $F_k=k$ the emission of $k$ photon bunches. At larger $E_J$, a more complex behavior is observed in quantitative agreement with numerical simulations. 
\end{abstract}

\pacs{74.50+r, 73.23Hk, 85.25Cp}
\date{\today, version 7 pm}

\maketitle
\section{Introduction}
The pioneering work of Max Planck on light emission by hot matter led to the recognition of the granular character of light and to the concept of photon. The quantum theory then explained how electrons in atoms occupy discrete energy eigenstates and how transitions between these states radiate single photons. The corresponding photon emission rate is governed by the light-electrical charge coupling strength, measured by the ratio between the vacuum impedance $Z_0$ and the quantum of resistance $R_\mathrm{K}=h/\mathrm{e}^2$, namely the fine structure constant $\alpha_\mathrm{QED} = Z_0 / 2 R_\mathrm{K} \simeq 1/137$ \cite{alpha}. The smallness of $\alpha_\mathrm{QED}$ places usual light emission in a perturbative regime of quantum electrodynamics (QED).

On the other hand, emission of photon multiplets (bunches with always the same number $k$ of photons) in a single event only occurs in special circumstances, as in atomic cascades, or in nonlinear media able to split single photons into several ones. For instance, parametric down-conversion is commonly used to produce pairs of strongly correlated photons, and emission of photon triplets has even been achieved \cite{PhysRevA.84.033823, chang2020observation}. Here, we consider multiphoton emission in the generic context of electrical circuits with a quantum coherent conductor steadily maintained out-of-equilibrium by a dc voltage source, and producing photons each time this conductor is traversed by a charge carrier \cite{ingold92,0803.0020,PhysRevLett.116.043602,PhysRevLett.106.217005}. The resulting QED of this type of circuits with designed light-matter coupling strength \cite{PhysRevB.91.205417,PhysRevB.93.075425, PhysRevB.95.125311,PhysRevX.6.031002,PhysRevLett.116.043602,JuhaNJP,PhysRevLett.110.267004, armour2013universal, PhysRevLett.111.247002} can provide e.g. sub-Poissonian photon sources 
\cite{PhysRevLett.86.700,PhysRevLett.93.096801,PhysRevB.81.155421,PhysRevB.81.115331, PhysRevB.92.195417, PhysRevX.9.021016, PhysRevLett.122.186804}, novel types of lasers \cite{cassidy2017,PhysRevLett.107.073901,PhysRevB.87.094511, PhysRevB.89.104502}, near-quantum limited amplifiers \cite{Jebari18,PhysRevApplied.11.034035}, squeezed radiation \cite{PhysRevLett.114.130403,PhysRevB.93.075425,PhysRevB.95.125311}, and interesting quantum state engineering resources \cite{ PhysRevB.102.155105, Peugeot2021, 1807.02364, AielloPhD}. 

\begin{figure}[!ht]
\begin{center}
\includegraphics[width =0.45\textwidth]{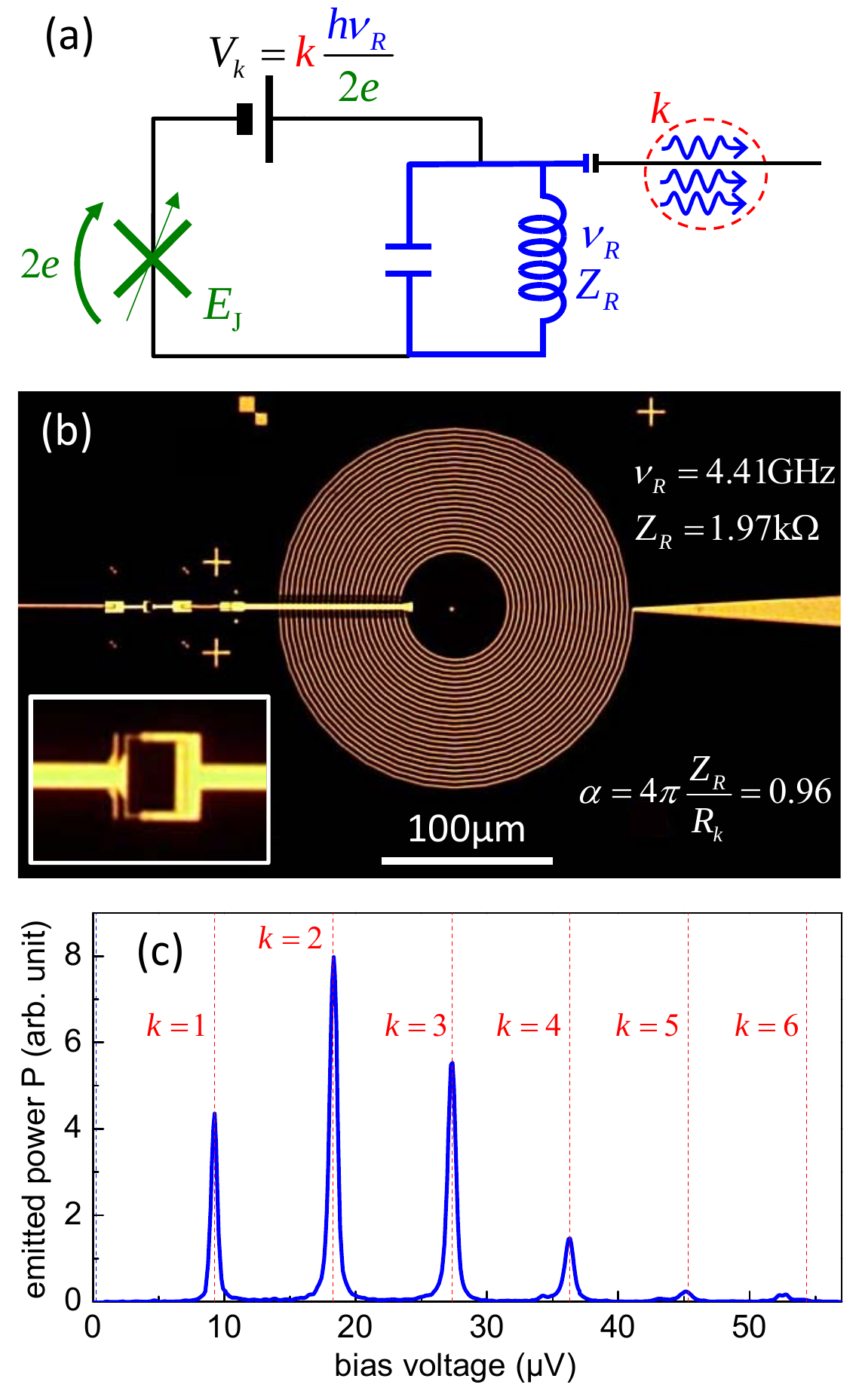}
\caption{\textbf{Principle of the experiment and first observation of multiphoton emission up to k=5 (run 1):} \textbf{(a)} A tunable Josephson junction (JJ) with energy $E_J$ (green cross) is connected in series with a dc voltage source $V$ and a microwave resonator (blue) of frequency $\nu_R$ and characteristic impedance $Z_R$. Current can flow only at certain values $V_k$ of $V$ for which the energy $2eV_k$ of a Cooper pair transferred across the circuit is entirely transformed into an integer number $k$ of photons in the resonator. Is the field leaking out of the resonator quantitatively understood and does it display k-photon bunches? \textbf{(b)} Optical micrograph of the sample showing a SQUID (magnetically tunable JJ - main picture and inset) connected to a high inductance coil (resonator). The electrical parameters are indicated and lead to a giant effective fine structure constant $\alpha \sim 1$. The bias and measuring lines are schematized in Fig. \ref{experimental_setup}.c of Appendix A. \textbf{(c)} Emitted power measured as a function of the bias voltage $V$ for a Josephson energy $E_J$ large enough to observe emission peaks up to $k = 5$. The black and red vertical dotted lines indicate the offset voltage and the $V_k$ values, respectively. Note that the small peak visible below $V_6$ is a spurious emission attributed to a high frequency mode of the circuit.}
\label{fig1}
\end{center}
\end{figure}

The particular circuit of this work and the principle of the experiment are shown in the schematics of Fig.~\ref{fig1}(a). A tunable Josephson junction (JJ) with Josephson energy $E_J$, biased at a dc voltage $V$, is placed in series with a microwave resonator of frequency $ \nu_R= \omega_R / 2 \pi $ and characteristic impedance $Z_R $. In addition, the resonator is capacitively overcoupled to a measurement line, into which photons leak at an energy decay rate $\kappa = \omega_R/Q$. By design, $Z_R $ is of the order of $R_K$, thus placing microwave emission in the non-perturbative regime far away from the conventional QED regime. Note that $V$ is kept much smaller than the superconducting gap voltage of the JJ electrodes, so that no electrons can tunnel through the JJ at low temperature. Only Cooper pairs with charge $-2\mathrm{e}$ can thus tunnel, provided that the energy $2\mathrm{e} V$ delivered by the source is entirely converted into an integer number $k$ of photons in the resonator. These inelastic processes occur only at particular bias values $V_k$ such that
\begin{equation}
\label{resonant} 
2\mathrm{e}V_k=k\, h \nu_R,\ k =1,2, 3...
\end{equation}
The aim of this experimental work is to obtain these k-photon bunches with a high brightness, to compare the photon fluxes to theoretical predictions, and to obtain a signature of the k-granularity.

\section{Theory}

The Hamiltonian of the circuit is the sum of the resonator Hamiltonian $\hbar \omega_R ( \hat{a}^\dagger \hat{a}+1/2)$, with $\hat{a}^{(\dagger)}$ the photon annihilation (creation) operator, and of the Josephson Hamiltonian $\hat{H}_J = -E_{\rm {J}} \cos \hat{\phi}_J$ with $\phi_J$ the superconducting phase difference across the JJ. The voltage source imposes a total phase difference across the circuit increasing linearly with time $t$, $\phi_V = \omega_\mathrm{J}t = \hat{\phi}_J + \hat{\phi}_R$, with $\nu_\mathrm{J} = \omega_\mathrm{J}/ 2 \pi = 2\mathrm{e} V /h $ the Josephson frequency, $ \hat{\phi}_R = \sqrt{\alpha}(\hat{a}^\dagger + \hat{a})$ the phase across the resonator, and $\alpha= 4\pi Z_R/R_k$. The time dependant Hamiltonian of the circuit is thus

\begin{equation}
\hat{H}= \hbar \omega_R \hat{a}^\dagger \hat{a} - E_{{\mathrm J}}\cos[\omega_\mathrm{J}t- \sqrt{\alpha}(\hat{a}^\dagger + \hat{a})],
\label{H_tdependent} 
\end{equation}
up to the resonator zero-point energy. Note that $\phi_J$ being conjugate to the number $N$ of Cooper pairs transferred through the JJ and $\hat{H}_J$ being the sum of the operators $e^{\pm i \hat{\phi_J}}$ that increase or decrease $N$ by one unit, $\hat{H}_J$ couples Cooper pair transfer to photonic excitations in the resonator \cite{PhysRevLett.106.217005, PhysRevLett.111.247002}. Hamiltonian (\ref{H_tdependent}) shows that the strength of this coupling is given by $\alpha$, which is the charge-radiation coupling constant \cite{ingold92} of our one-mode circuit, and plays the same role as the fine structure constant $\alpha_\mathrm{QED}$ in atomic physics.
This coupling results in inelastic Cooper pair tunneling and in a dc current flowing through the circuit in the vicinity of voltages $V_k$. At $\omega _J = k \omega_R + \delta_k $, the effective Hamiltonian obtained within the rotating-wave approximation (RWA) takes the form \cite{PhysRevB.92.054508,PhysRevB.93.041418(R)} 
\begin{equation}
\hat{H}_k= - \frac {E_J e^{- \frac{\alpha}{2}}} {2} \alpha^{\frac {k} {2}} \left[ e^{- i \delta_k t} \hat{B}_k (i\hat{a}^\dagger)^k + \mathrm{h.c.} \right],
\label{H_rwa} 
\end{equation}
where $\mathrm{h.c.}$ denotes Hermitian conjugation and
\begin{equation}
\hat{B}_k = \sum_{n=0}^{\infty}\frac{n!}{(n+k)!}\;L_n ^{(k)} (\alpha) \ket{n} \bra{n}
\label{Bk}
\end{equation}
is a diagonal operator in the Fock state basis $\{\ket{n}\}$ involving the generalized Laguerre polynomials $L_n ^{(k)} (\alpha)$ \cite{PhysRevB.93.041418(R)}. The Cooper pair translation operators $e^{\pm i \hat{\phi_\mathrm{J}}}$ have thus been transformed into creation and annihilation operators $\hat{a}^{(\dagger)k}$ adding or removing bunches of $k$ photons to/from the resonator. Under a constant voltage, a steady state situation is reached, characterized by an average number of photons in the resonator, the occupation number $\left<n\right>$ : Cooper pairs tunnel across the JJ at a rate $\gamma_k$ and produce photons in the resonator; these photons leak at an average rate $\Gamma_k= k \gamma_k = \kappa \left<n\right>$ in the $a_{out}$ modes of the measurement line (see Fig.~\ref{experimental_setup} in Appendix A). This photon rate is expected to be proportional to the square of the prefactor $E_J e^{-\alpha /2} {\alpha}^{k /2}/2$ in Hamiltonian (\ref{H_rwa}), and to also depend, through $\hat{B}_k$, on the actual photon distribution probability inside the resonator.

In the limit of vanishing $E_J$ the situation is simpler since the resonator has time to empty before a new Cooper pair tunnels and a new bunch of $k$ photons is emitted. The rate
\begin{equation}
\gamma_k = \frac{\Gamma_k}{k} = -\left(\frac{E_J}{\hbar \omega_R}\right)^2 \frac{\alpha^k e^{-\alpha}}{k\;k!} \frac{ Q \omega_R }{1+\left(\frac{2Q\delta_k}{k\omega_R}\right)^2}
\label{gamma-of-k} 
\end{equation}
is obtained in this case from a standard calculation of the Purcell relaxation rate for Hamiltonian (\ref{H_rwa}), which holds when $\gamma_k\ll\kappa$, and coincides with the prediction of Dynamical Coulomb Blockade theory \cite{PhysRevLett.122.186804}. In this regime of well separated tunnel events, the microwave radiation consists of separated bunches of $k$ photons. This granularity of the energy flow is naturally measured by the photon Fano factor $F_k$, defined as the ratio of the variance to the mean number $\Gamma_k t$ of emitted photons during a time $t>\Gamma_k^{-1}$. Assuming a Poissonian electrical current with a Cooper pair Fano factor of 1, and $k$ photons per Cooper pair crossing the circuit, one predicts $F_k = k$. Now, in the microwave domain, no wideband photon counters exist and the emitted photons cannot be counted during a given period of time. Instead, the field statistics can be characterized by the normalized second order (intensity-intensity) correlation function
\begin{equation}
 g^{(2)}(\tau) = \frac{\left<\hat{a}_\mathrm{out}^\dagger(0)\hat{a}_\mathrm{out}^\dagger(\tau)\hat{a}_\mathrm{out}(\tau)\hat{a}_\mathrm{out}(0)\right>}{\left<\hat{a}_\mathrm{out}^{\dagger}\hat{a}_\mathrm{out}\right>^2},
 \label{g2} 
\end{equation}
which can be interpreted as the probability for two photons separated by a delay $\tau$ to leak in the same electromagnetic mode. Then, $F_k$ can be computed from $g^{(2)}$ \cite{g2ToFano} as
\begin{equation}
 F_k = 1+2\Gamma_k\int_{0}^{+\infty}{\left[g^{(2)}(\tau)-1\right]\mathrm{d}\tau}.
\end{equation}

In the strong brightness regime at large $E_J$, the resonator population feeds back to the emission dynamics \cite{Armour2017}. A numerical integration of the evolution of the system under the influence of Hamiltonian (\ref{H_rwa}) and radiative losses with rate $\kappa$ is necessary to predict both the emission rate $\Gamma_k$ and the Fano factor $F_k$.

\section{Implementation}
The $\alpha^k$ factor in Eq. (\ref{gamma-of-k}) calls for a large value of $\alpha$ to favor the multiphoton emission beyond the already observed k=2 case \cite{PhysRevLett.106.217005}. Standard on-chip microwave resonator designs yield a characteristic impedance smaller than the vacuum impedance $Z_V \simeq 377 \, \Omega$, with typically $\alpha \sim0.05$. To approach $\alpha \sim1$, we use a spiral coil resonator [see Fig.~\ref{fig1}(b)] etched in a 150~nm-thick niobium film sputtered onto a quartz substrate (low dielectric constant $\epsilon_r \simeq 3.8$), whereas the JJ is a superconducting quantum interference device (SQUID) with an $E_J$ of a few µeV magnetically tunable down to almost zero. The resonator capacitance being the spurious capacitance to ground of its coil in parallel with the JJ capacitance, its resulting central frequency and characteristic impedance are $\nu_R = $4406~MHz and $Z_R = 1.97$~k$\Omega$, which corresponds to $\alpha=0.96$.

The data reported here were collected over three different experimental runs by measuring a sample  previously used to demonstrate photon antibunching at k=1 \cite{PhysRevLett.122.186804}. Over these runs that extended over four years, the tunnel resistance of the SQUID increased from 220 to 330 $ \rm {k} \Omega$ due to aging, leading to a similar decrease of the maximum reachable $E_J$ value. Runs 1 and 2 were performed in a dilution refrigerator (DR) with a liquid helium cryostat, whereas run 3 used a cryo-free DR with a pulse tube. As a result, the bias voltage noise was $\sim4$ nV  in run 2 and $\sim80$ nV in run 3, which corresponds to Josephson frequency noises with standard deviations of about 2~MHz and 38~MHz, respectively (see Appendix F). For all runs, the sample was placed in the same shielded sample-holder, and was thermally anchored to the mixing chamber of the DR. The sample was connected through the same bias tee to a dc line with a filtered voltage divider, and to a 50~$\Omega$ microwave detection circuit. With such a low impedance detection scheme, the quality factor $Q$ of the resonator cannot be precisely controlled and has to be measured precisely in-situ (see Appendix C - $Q=36.6$ and $72$ in run 2 and 3). The detection line is made of a 90$^\circ$ hybrid coupler acting as a microwave beam splitter towards two nominally identical amplified lines $1$ and $2$ (see Appendix A). At room temperature, the signals $v_1(t)$ and $v_2(t)$ from the two lines were measured in different ways: in run 1 their powers are measured using two quadratic detectors, whereas in runs 2-3, they are bandpass filtered and down-converted to the 0-625~MHz frequency band using two mixers sharing the same local oscillator at $\nu_{\mathrm{LO}}= 4.71$~GHz. In this later case, the two output quadratures are then digitized at 1.25~GSamples/s. The relevant correlation functions are then computed to obtain the emitted power spectral density, the total emitted power, the second order coherence function $g^{(2)}(\tau)$, and the corresponding Fano factor $F_k$ at the output of the resonator. As in \cite{PhysRevLett.122.186804}, this two-line measurement setup \`a la Hanbury-Brown and Twiss is a convenient way to remove the contribution of the technical noises from the determined quantities (see Appendix B).

\section{Measured and simulated emitted power}
In the first experiment (run 1) we simply sweep the voltage V and integrate the received power over a bandwidth larger than the resonator one. We observe regularly spaced peaks [see Fig. \ref{fig1}(c)] that correspond to the k-photon excitation mechanism described above at $V=V_k$, for $k=1$ to $5$. In order to observe emission at such large values of $k$, we used the external magnetic field to tune our system to a large Josephson energy $E_J/h \nu_R \sim 0.1-0.2 $ (which could not be determined precisely due to the hysteretic magnetic behavior of the sample - See Appendix E). One notices in particular that the $k=2$ and $k=3$ peaks are stronger than the $k = 1$ peak, a situation that does not naturally occur in atomic physics because of the smallness of the fine structure constant. The last visible peak that appears below the voltage expected for $k=6$ does not correspond to a multiple order resonance, and results from a spurious mode of the setup that could be strongly reduced in run 2.

\begin{figure}[!ht]
\begin{center}
\includegraphics[width =0.45\textwidth]{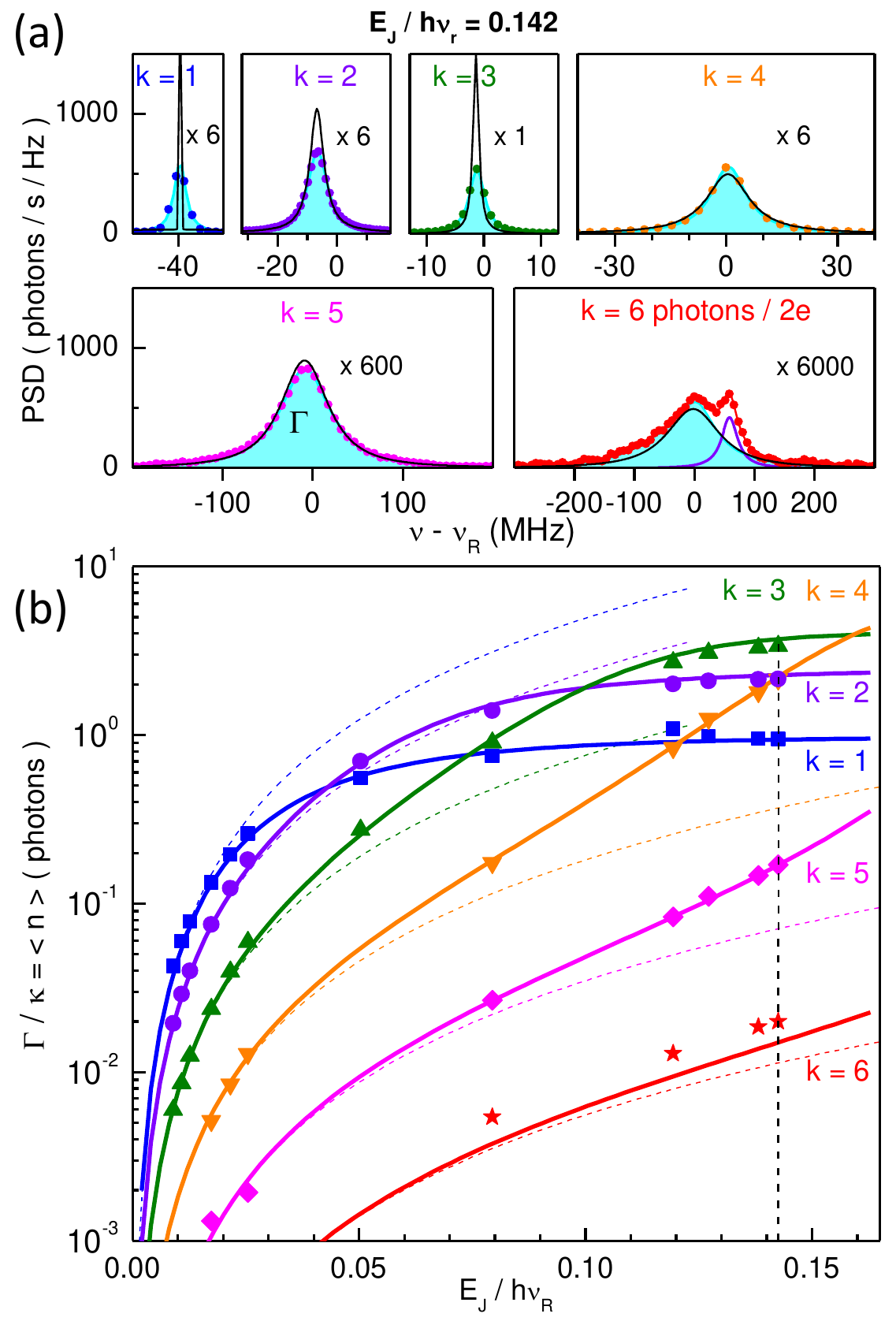}
\caption { \textbf{Multiphoton emission spectra and rates (run 2):}
Measured and simulated emitted power for $k=1$ to 6. \textbf{(a)} Examples of measured (dots) power spectral densities (PSD) taken at $ E_J / h \nu_R \sim 0.142 $ around the resonator frequency $\nu_R$, for different bias voltages $V_k+\delta V_k$ corresponding to residual frequency detunings $\delta_k/2\pi = 2\mathrm{e}\delta V_{k} /h k =$ -39.4, -6.0, -1.2, 1.4, -7.5, and 0.8 MHz for $k=1,..,6$. A vertical magnification factor with respect to the left axis scale is indicated for each peak. Cyan filled peaks are Lorentzian fits of the PSDs (for k = 6, a two-lorentzian fit gives a second spurious mode in violet). The cyan areas are the photon rates $ \Gamma_k$ used for comparison with simulations in panel (b). Bare simulations of the discretized spectral densities at zero detuning (black solid lines) are shown (horizontally shifted) for comparison. \textbf{(b)} Measured (dots) and simulated (solid lines) reduced emission rates $\left<n\right>= \Gamma_k / \kappa$ for $k=1\, \rm{to} \, \, 6$ and twelve different $ E_J $ values. Dashed curves represent the number of photons obtained from the Purcell rate $\gamma_k$ of departure from the vacuum state (see text). Due to magnetic flux jumps in the SQUID, experimental $ E_J $ values were not precisely known and were fitted to minimize the difference between simulation and experiment in Log scale (see text and Fig. \ref{hysteresis} in Appendix E). No voltage noise is included in the simulation and no vertical scaling of the data is applied after calibration. The +/- 5 \% systematic relative uncertainty on calibration plus the uncertainty on $k$ is about the symbol size. The vertical dashed line corresponds to the dataset in panel (a).}

\label{fig2}
\end{center}
\end{figure}

 In order to analyze more in-depth the multiphoton emission process observed, the power spectral densities (PSD) of the emitted radiation were measured in run 2, close to each $V_k$ up to $k = 6$, and as a function of $E_J$ in a range $ E_J / h \nu_R = 0.01 -0.142$ (see Fig. \ref{fig2}). Spectra for the highest $E_J$ are displayed in panels (a) and show a maximum brightness for $k=3$. The applied bias voltages being slightly offset by $\delta V_k$, the emission peaks are frequency shifted from the resonator frequency $\nu_R$ by $\delta_k/2\pi = 2 \text{e}\delta V_k/h k$. The spectral shapes are well fitted by Lorentzian peaks up to $k=5$. The $k=6$ peak is now clearly visible, but with a small parasitic contribution on its high-frequency side (which could be due to the emission of one photon in the resonator and another one in a 22.1~GHz spurious mode of the circuit).
 
 We then compare the measured spectra with those predicted from simulations of the master equation of the circuit with Hamiltonian (\ref{H_rwa}) and rate $\kappa$, without including the $\sim$~2~MHz Josephson frequency noise nor the detunings $\delta_k$. These theoretical spectra are $\delta_k$-shifted and superposed to the measured ones in Fig.  \ref{fig2}(a). At k=1 the theoretical spectrum is close to a monochromatic line at the Josephson frequency, so that the width of the measured spectrum is almost entirely due to the bias voltage noise. At higher $k$, energy conservation only imposes the sum of the $k$ photon frequencies to be equal to the Josephson frequency, so that the emission width is wider. The simulated spectra have widths that vary with $k$ in a non-trivial way, and reproduce reasonably well the measured ones. The experimental emission peaks are also slightly widened by the noise. 
 
 The integral of the Lorentzian fit of a PSD yields the corresponding total photon flux $\Gamma_k$ leaking out of the resonator. Note that the measured flux gives the resonator occupation $\left<n_k\right> = \Gamma_k/\kappa$ because the resonator population decays dominantly by the photon emission into the measurement line. These photon fluxes are displayed as dots in the panel (b) of Fig. \ref{fig2}, for twelve different values of the coil current and consequently of $E_J$. Unfortunately, the magnetic flux through the SQUID being hysteretic when ramping the coil current (see Appendix E), we had to use the following procedure to determine these $E_J$ values and compare the data to simulations: the six photon fluxes $\Gamma_k$ are simulated over the full $ E_J / h \nu_R = 0.01 -0.15$ range. The simulated curves are then used to fit the whole dataset with the twelve $E_J$ values as fitting parameters. These fitted $E_J$ are then validated by the independently characterized hysteretic behavior of the SQUID (see Appendix E). The simulated and measured $\Gamma_k$ match quantitatively over more than three orders of magnitude, with however a 25-30 percent discrepancy for $k=6$. The theoretical predictions of Eq.(\ref{gamma-of-k}) in the weak brightness regime are also shown as dotted lines. They also account for the data at small $E_J$ and occupation $\left<n\right>$, but fail to predict the brightness increase when the resonator occupation differs significantly from zero and feeds back on the emission dynamics.

\section{Granularity of the microwave emission}
The microwave granularity was measured in run 3 for $k=$1 to 4, by computing the $g^{(2)}$ functions from the digitized demodulated signals $v_1$ and $v_2$ (see Appendix B), for six different values of $E_J$ (see Fig. \ref{g2set} in Appendix G), an example being also shown in Fig. \ref{fig3}(b). Due to the magnetic hysteresis of the SQUID already mentioned, these $E_J$ values and their uncertainties are now inferred from the comparison of the measured average photon number $\left<n\right>$ with simulations. The large $\sim~80$~nV voltage noise observed in run 3 has now to be included in the simulation (see Appendix H) for a quantitative agreement. The corresponding Fano factors are shown in Fig. \ref{fig3}(c) and compare reasonably well with numerical simulations. As expected, the simulated $F_k$ tend towards $k$ at vanishing brightness but depart significantly from it when $E_J$ is increased. At the lowest $E_J$ compatible with a reasonable measuring time ($\sim$~72 hours), we measure $F_{1,2,3,4} = 0.7\pm 0.1$, $1.8\pm0.1$, $3.5\pm0.3$, and $4.5\pm0.6$ respectively, close to the expected bunch sizes k=1, 2, 3, and 4. This result, together with the quantitative understanding of the total emitted power, are the main results of the present work. This $k$ granularity of the emission at low $E_J$ means that the resonator is prepared in Fock state $ \ket{k}$, at each Cooper pair tunneling event. Note that other circuit-QED devices can prepare such Fock states, but in a series of about k successive operations involving a superconducting qubit coupled resonantly \cite{HofheinzMartinis} or dispersively \cite{Krastanov} to a resonator. All these devices can thus  be regarded as sources of k-photons multiplets.

\begin{figure}[h!]
\begin{center}
\includegraphics[width =0.48\textwidth ]{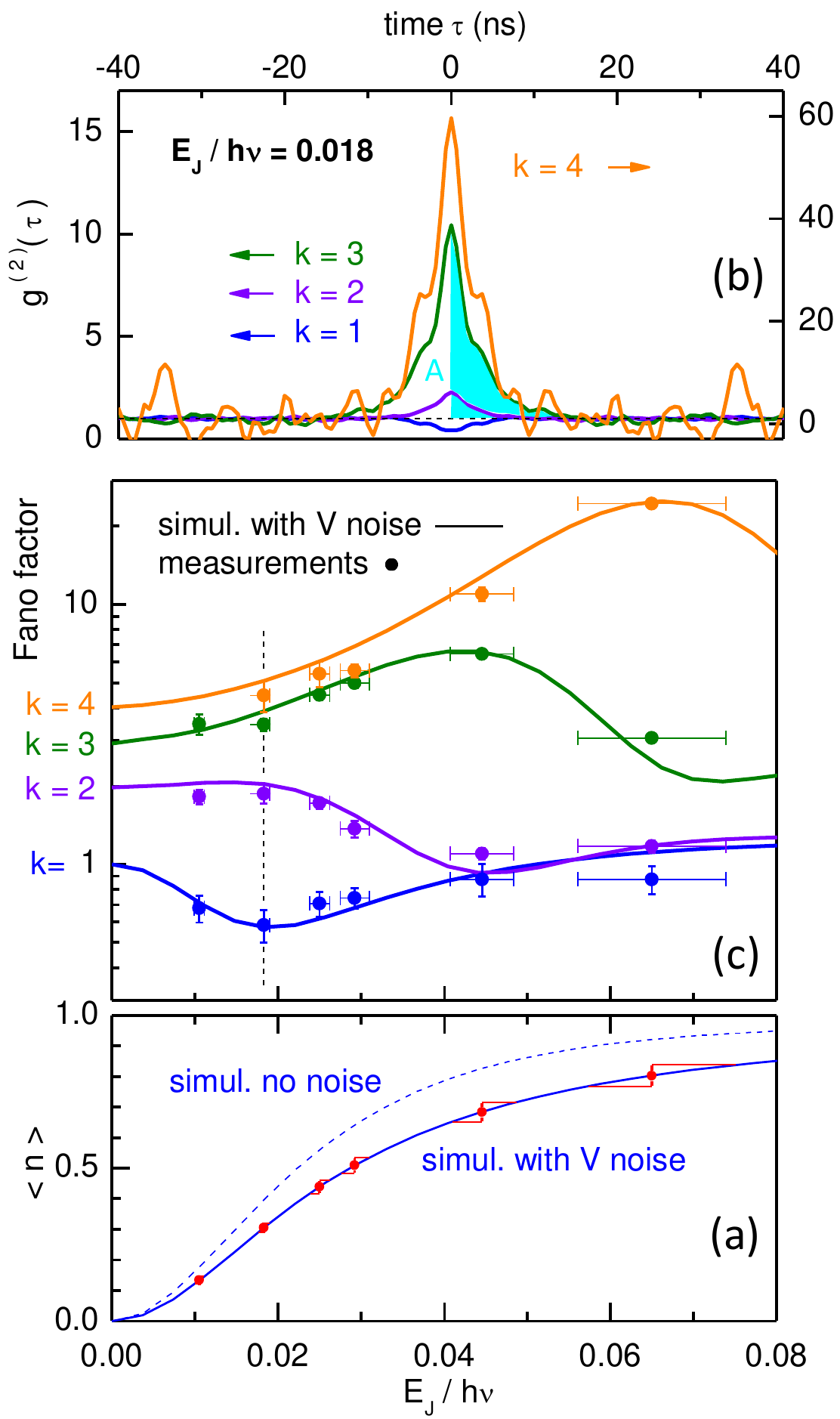}

\caption{\textbf{Noise statistics and Fano factors of the leaking microwave (run 3)}. \textbf{(a)} Calibration of the $E_J$ values by placing the measured $\left<n\right>$ (dots) on a simulation of the $\left<n\right>(E_J)$ curve including the effect of the observed voltage noise (solid line). Simulation with a noiseless bias voltage (dashed line) is also shown for comparison. The uncertainty on $\langle n \rangle$ (vertical error bars) yields the uncertainty on $E_J$ (horizontal error bars). \textbf{(b)} Example of a set of $g^{(2)}$ functions measured at a small $E_J / \hbar \omega_R = 0.018$. The area between $g^{(2)}$ and $1$, integrated over the $0-40$~ns time interval (shown in cyan for k=3) enters the Fano factor of panel (c). \textbf{(c)} Experimental Fano factors (dots) and simulated ones including voltage noise (solid curves). Horizontal error bars correspond to those of panel (a), whereas vertical ones correspond to $\pm2$ standard deviations. The vertical dashed line corresponds to the raw data of panel (a).
}
\label{fig3}
\end{center}
\end{figure}

Upon increasing $E_J$, the Fano factors show a complex behavior with a dip for $k=1$ and a peak-dip structure for $k>1$ [see panel (c)], although the increase in the resonator occupation $\left<n\right> \lesssim 1$ remains moderate [see Fig. \ref{fig3}.(a)]. In essence (see Appendix I), these variations are a consequence of two competing nonlinear effects. First, at moderate occupation, the $k$-parametric term $\sim (\hat{a}^\dagger)^k+\hat{a}^k$ of Hamiltonian (\ref{H_rwa}) has matrix elements increasing rapidly with the number $n$ of photons. This results in an emission stimulated by the photons already present in the resonator, an enhanced emission rate, and a superlinear Fano factor $F_k>k$ that indicates an additional bunching of the k photon-multiplets \cite{PhysRevLett.111.247002, armour2013universal, PhysRevB.86.054514}. Second however, at larger $E_J$ and occupation $\left<n\right>$, the Josephson nonlinearity encoded in $\hat{B}_k$ reduces strongly the drive strength, at different occupation levels that depend on $k$. For $k=1$, this last effect even leads to a strong antibunching (see \cite{PhysRevLett.122.186804} and $g^{(2)}(0)<1$ in Fig. \ref{g2set} of Appendix G).

Finally, beyond the k-granularity of the photon emission, an interesting point to note is the quantum nature of the resonator field $\hat a$ and radiated fields $\hat a_{out}$: as the emission is similar to a k-photon spontaneous parametric down-conversion [see the $\hat{B}_k$ modified $(\hat{a}^\dagger)^k+\hat{a}^k$ term in Hamiltonian (\ref{H_rwa})], the Wigner functions of the fields are expected to be non-Gaussian and squeezed with a k-fold symmetry, as measured recently in \cite{chang2020observation} for $k=$3. Our experiment was designed to measure photon statistics and does not permit measurement of these Wigner functions. However we show an example of simulated ones in Appendix J to motivate their measurement in the future.

\section{Conclusion}

As a conclusion, our work shows that a high-impedance resonator in series with a voltage-biased Josephson junction produces, at particular voltages $V_k$, bunches of $k=1 ...,6$ photons per Cooper pair tunneling across the Josephson junction, with a brightness quantitatively understood. By measuring both the emitted power and the Fano factor of the system we have shown that the photons are indeed emitted as multiplets in a single event. Note that a similar emission process was also recently observed in a richer multimode environment \cite{AielloPhD}. Our simple system provides an interesting test bench for quantum optics experiments in the strong charge-radiation coupling regime provided by the high impedance moreover in a steady-state out-of-equilibrium situation imposed by the voltage source, a regime far from atomic physics. Beyond illustrating and clarifying a new regime of quantum optics, such simple  photon sources complete the quantum microwaves toolbox.\\

\noindent
\begin{tabular}[t]{ll} 
\includegraphics[width =0.14\textwidth, valign=c] {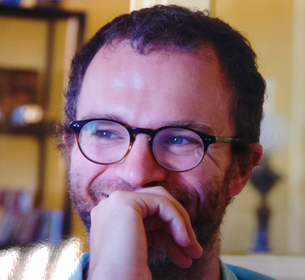} & \parbox{59mm}{\textit{This article closes a series of works on microwave quantum optics with voltage-biased Josephson circuits, inspired by our passionate colleague and friend Fabien Portier, who passed away on December 2, 2020.}}\\
\end{tabular}

\section*{Acknowledgement}{
We gratefully acknowledge stimulating discussions within the Quantronics and Nanoelectronics groups and with K. Mölmer and A. Armour, as well as the technical help from P. Jacques. This work
received funding from the European Research Council (Horizon 2020/ERC Grant Agreement
No. 639039 and NSECPROBE), from the French LabEx PALM (ANR-10-LABX-0039-PALM), from the French ANR (contracts GEARED ANR-14-CE26-0018-01 and SIMCIRCUIT ANR-18-CE47-0014-01), from the German-French ANR/DFG Grant JosephCharli, from IQST and the German Science Foundation (AN336/13-1).
}

\section*{Appendix A: Complete experimental setup}
\label{setup}

\begin{figure}[h]
\begin{center}
    \includegraphics[width = 0.45\textwidth]{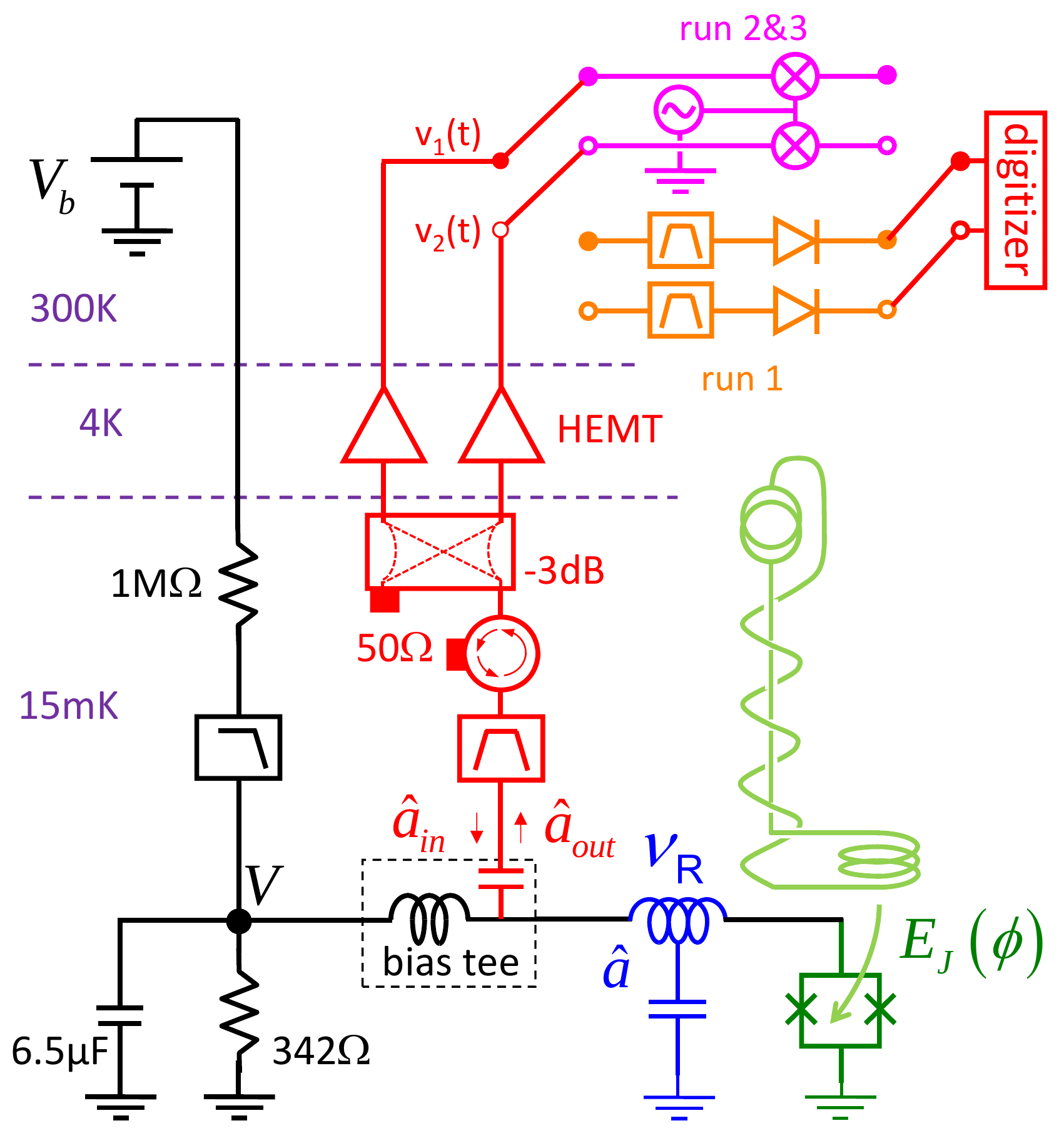}
    \caption{\textbf{Experimental setup:} Schematics of the whole measurement circuit used in the different experimental runs (see text).}
    \label{experimental_setup}
\end{center}
\end{figure}

Figure \ref{experimental_setup} presents the electrical schematics of our experimental bias and detection circuits. As explained in the main text, our sample is anchored at the bottom plate of a dilution refrigerator and connected to the measurement chain. The black sub-circuit with a bias-tee is used to apply a finite bias to the junction (in dark green), which generates the emission of photons into the resonator (in blue). An external magnetic coil (in light green) threads a magnetic flux $\phi$ through the SQUID to tune its Josephson energy $E_\mathrm{J}$. The signal leaking from the resonator (mode $a_{out}$) goes through a band-pass filter and a circulator before being split over two lines in a Hanbury-Brown and Twiss-like setup and then amplified (in red). Depending on the experimental run, we either send each signal to a band-pass filter and a diode (in orange, run 1), or heterodyne them (in purple, runs 2 and 3), before digitizing them.

\section*{Appendix B: Determination of spectral densities, total powers and $g^{(2)}$ functions (run 2 and 3)}
\subsection{Model for the detection chain}

We describe here the same procedure as the one given in the supplemental materials of \cite{PhysRevLett.122.186804}.

The input-output formalism links the resonator operator $\hat{a}$ to the in-going and out-going transmission line operators $\hat{a}_{\mathrm{in}}$, $\hat{a}_{\mathrm{out}}$ by $\sqrt{\kappa}\hat{a}(t) = \hat{a}_{\mathrm{in}}(t)+\hat{a}_{\mathrm{out}}(t)$, with $\kappa = \omega_R/Q$ the energy leak rate. In our experimental setup, $\hat{a}_{\mathrm{in}}$ describes the thermal radiation coming from the $50~\Omega$ load on the isolator closest to the sample. Because this load is thermalized at $15~mK\ll \hbar\omega_R/k_B$, the modes impinging onto the resonator  can be considered in the ground state and the contribution of $ \hat{a}_{\mathrm{in}}$ to all the correlation functions vanishes. We thus take $\hat{a}_{\mathrm{out}}$ to be an exact image of $\hat{a}$, and all their normalized correlation functions as being equal.

As described in the previous section, the emitted signal is split between two detection chains, filtered, amplified, and mixed with a local oscillator before digitization. The beam-splitter is implemented as a hybrid coupler with a cold $50~\Omega$ load on its fourth port, and acts as an "amplifier" of gain $1/2$, adding to the signal a noise mode $\hat{h}_{bs}$ in the vacuum state. The different amplifying stages are summed up into one effective amplifier for each channel, with respective noise temperatures $T_\mathrm{N}^{(1)} = 13.5$~K and $T_\mathrm{N}^{(2)}= 14.1$~K for runs 1 and 2 and $T_\mathrm{N}^{(1)} = 8.12$~K and $T_\mathrm{N}^{(2)}= 6.16$~K for run 3. The mixer used for heterodyning the signals also adds to them at least the vacuum noise.

The last step is the linear detection of the voltage $v_i(t)$ on channel $i$ by the acquisition card. It is classical and does not have to be modelled at the quantum level. After digitization, we process chunks of signal of length 1024 samples to compute the analytical signal $S_i(t) = v_i(t)+H(v_i)(t)$, where $H$ is here the discrete Hilbert transform. Because computing the analytical signal  $S_i(t)$ from the heterodyned $v_i(t)$ is equivalent to measuring its two quadratures, which are non-commuting observables, quantum mechanics imposes an additional noise mode. We sum this digitization noise and the  heterodyned noise into a single demodulation noise.

Ultimately, we record measurements of $\hat{S}_1(t)$ and $\hat{S}_2(t)$, with $\hat{S}_i \propto \hat{a}_i+\hat{h}_i^{\dagger}$ and $\hat{h}_i$ an effective thermal noise summing the different detection step contributions (in practice, the dominant noise contribution stems from the amplifiers closest to the sample).  Note that $\hat{a}_i(t)\propto \hat{a}_\mathrm{out}(t-\tau_i)$ with $\tau_i$ the propagation time on channel $i$, and that the signal model described above is valid only within the bandpass of the filters.

\subsection{Computing correlations (run 2 and 3)}

For each chunk of signal recorded on line $i$ we compute a chunk of $S_i(t)$ of the same length 1024. We then compute the desired correlation functions as
\begin{equation*}
    C_{X,Y}(\tau)=\left<X^*(t)Y(t+\tau)\right> = F^{-1}(F(X)^*F(Y))
\end{equation*}
where $\left<\cdots\right>$ stands for the average over the length of the chunk and $F$ is the discrete Fourier transform. Finally we average the correlation functions from all the chunks and store this result for further post-processing.

To illustrate how we reconstruct the information on $\hat{a}$ from $S_1$ and $S_2$, let's consider the first order coherence function $g^{(1)}(\tau) = \left<\hat{a}^{\dagger}(t)\hat{a}(t+\tau)\right>/\left<\hat{a}^{\dagger}\hat{a}\right>$. We start with the product
\begin{equation*}
\begin{matrix}
    S^*(t)S(t+\tau) & \propto & \hat{a}^{\dagger}(t)\hat{a}(t+\tau)+ \hat{h}(t)\hat{h}^{\dagger}(t+\tau)\\ 
    & &+\hat{a}^\dagger(t) \hat{h}^{\dagger}(t+\tau) + \hat{h}(t)\hat{a}(t+\tau)
\end{matrix}
\end{equation*}
on a single line, i.e. with $S$ being either $S_1$ or $S_2$.
Then, we consider that the noise added by an amplifier cannot be affected by the state of the resonator, so that $\hat{a}$ and $\hat{h}$ are independent and thus uncorrelated, hence
\begin{equation*}
    \left<\hat{a}(\tau)\hat{h}(t+\tau)\right> = \left<\hat{a}(\tau)\right>\left<\hat{h}(t+\tau)\right> = 0,
\end{equation*}
as there is no phase coherence in the thermal noise ($\left<\hat{h}\right> = 0$). We thus have
\begin{equation*}
\left<S^{*}(t)S(t+\tau)\right> \propto \left<\hat{a}^{\dagger}(t)\hat{a}(t+\tau)\right>+\left<\hat{h}(t)\hat{h}^\dagger(t+\tau)\right>.
\end{equation*}
Hence, at zero bias voltage $V=0$ (the so-called \textit{off} configuration),
\begin{equation*}
\left<S^*(t)S(t+\tau)\right>_\mathrm{off}\propto \left<\hat{h}(t)\hat{h}^{\dagger}(t+\tau)\right>,
\end{equation*}
whereas at the finite voltage $V$ for the multiphoton emission (so-called \textit{on} configuration)
\begin{equation*}
    \left<S^*(t)S(t+\tau)\right>_\mathrm{on}\propto\left<\hat{a}^\dagger(t)\hat{a}(t+\tau)\right>+\left<S^*(t)S(t+\tau)\right>_\mathrm{off}.
\end{equation*}
Consequently
\begin{equation*}
    g^{(1)}(\tau) = \frac{\left<S^*(t)S(t+\tau)\right>_\mathrm{on}-\left<S^*(t)S(t+\tau)\right>_\mathrm{off}}{\left<S^*S\right>_\mathrm{on}-\left<S^*S\right>_\mathrm{off}}.
\end{equation*}
Now, as we are considering states of the resonator with at most a few photons, we typically have $\left<S^*S\right>_\mathrm{off}\simeq\left<S^*S\right>_\mathrm{on}\gg\left<S^*S\right>_\mathrm{on}-\left<S^*S\right>_\mathrm{off}$.
From there, any small fluctuation of the gain of the detection chain or of the noise temperature during the experiment reduces greatly the contrast on $g^{(1)}(\tau)$. We thus rely on the cross-correlation $X(\tau) = \left<S_1^*(t)S_2(t+\tau)\right>$ on the two lines rather than on the previous auto-correlation on one of them. (Note that due to a residual cross-talk between the two channels this cross-correlation averages to a finite value even in the \textit{off} position, but which is 60 dB lower than the auto-correlation of each channel). We hence use
\begin{equation*}
    g^{(1)}(\tau) = \frac{X(\tau)_\mathrm{on}-X(\tau)_\mathrm{off}}{X(0)_\mathrm{on}-X(0)_\mathrm{off}}.
\end{equation*}

A similar treatment allows to compute $g^{(2)}(\tau)$ with slightly more complex calculations. The classical Hanburry-Brown-Twiss experiment correlates the signal power over the two channels, i.e. extracts $g^{(2)}(\tau)$ from $\left<S_1^*S_1(t)S_2^*S_2(t+\tau)\right>$. The \textit{off} value of this correlator is once again much bigger than the relevant information of the \textit{on-off} part, and any drift of the amplifiers would blur the averaged value of $g^{(2)}(\tau)$. This is why we use $C(t) = S_1^*(t)S_2(t)$ instead of $S_i^*(t)S_i(t)$ as a measure of the instantaneous power emitted by the sample, provided that the time delay between the two detection lines is calibrated and compensated for. We then have
\begin{equation}
 \begin{multlined}
        g^{(2)}(\tau) = \frac{\left<C(t)C(t+\tau)\right>_\mathrm{on}-\left<C(t)C(t+\tau)\right>_\mathrm{off}}{\left(\left<C\right>_\mathrm{on}-\left<C\right>_\mathrm{off}\right)^2}\\
        -2\frac{\left<C\right>_\mathrm{off}}{\left<C\right>_\mathrm{on}-\left<C\right>_\mathrm{off}}\\
        -\frac{\left[X(\tau)_\mathrm{on}-X(\tau)_\mathrm{off}\right]X(-\tau)_\mathrm{off}}{\left(\left<C\right>_\mathrm{on}-\left<C\right>_{\mathrm{off}}\right)^2}\\
        -\frac{\left[X(-\tau)_\mathrm{on}-X(-\tau)_\mathrm{off}\right]X(\tau)_\mathrm{off}}{\left(\left<C\right>_\mathrm{on}-\left<C\right>_\mathrm{off}\right)^2},
 \end{multlined}
\end{equation}
where $\left<C\right>$ stands for $\left<C(t)\right>$.

\subsection{Computing power spectral densities, total emitted powers and Fano factors from correlations}

From the raw cross correlation $C(\nu)$ (Fourier transform of $C(t)$), we compute the normalized emitted power spectral density
\begin{equation*}
\mathrm{PSD}(\nu)=\frac{C(\nu)_\mathrm{ON}-C(\nu)_\mathrm{OFF}}{C(\nu)_\mathrm{OFF}.}
\end{equation*}

The photon emission rate
\begin{equation*}
    \Gamma = \frac{k_\mathrm{B}T_\mathrm{N}}{h}\int_{\mathrm{BW}}{\frac{\mathrm{PSD}(\nu)}{\nu}\mathrm{d}\nu},
\end{equation*}
is then obtained by integrating the PSD over the bandwidth (BW) of the resonator, with $T_\mathrm{N} \sim 7.5~K$ the effective noise temperature of the cross correlated signal. We then use this photon emission rate $\Gamma$ and the fully corrected $g^{(2)}(\tau)$ described above to compute the Fano factors
\begin{equation*}
    F_k=1+2\Gamma_k\int_{0}^{+\infty}\left[1-g^{(2)}(\tau)\right]d\tau.
\end{equation*}
The experimental $g^{(2)}(\tau)$ and the estimation of the error on $F_k$ are presented in Appendix G.

\section*{Appendix C: Resonator parameter determination}
\label{section_resonator}
\begin{figure}[H]
\begin{center}
    \includegraphics[width = 0.45\textwidth]{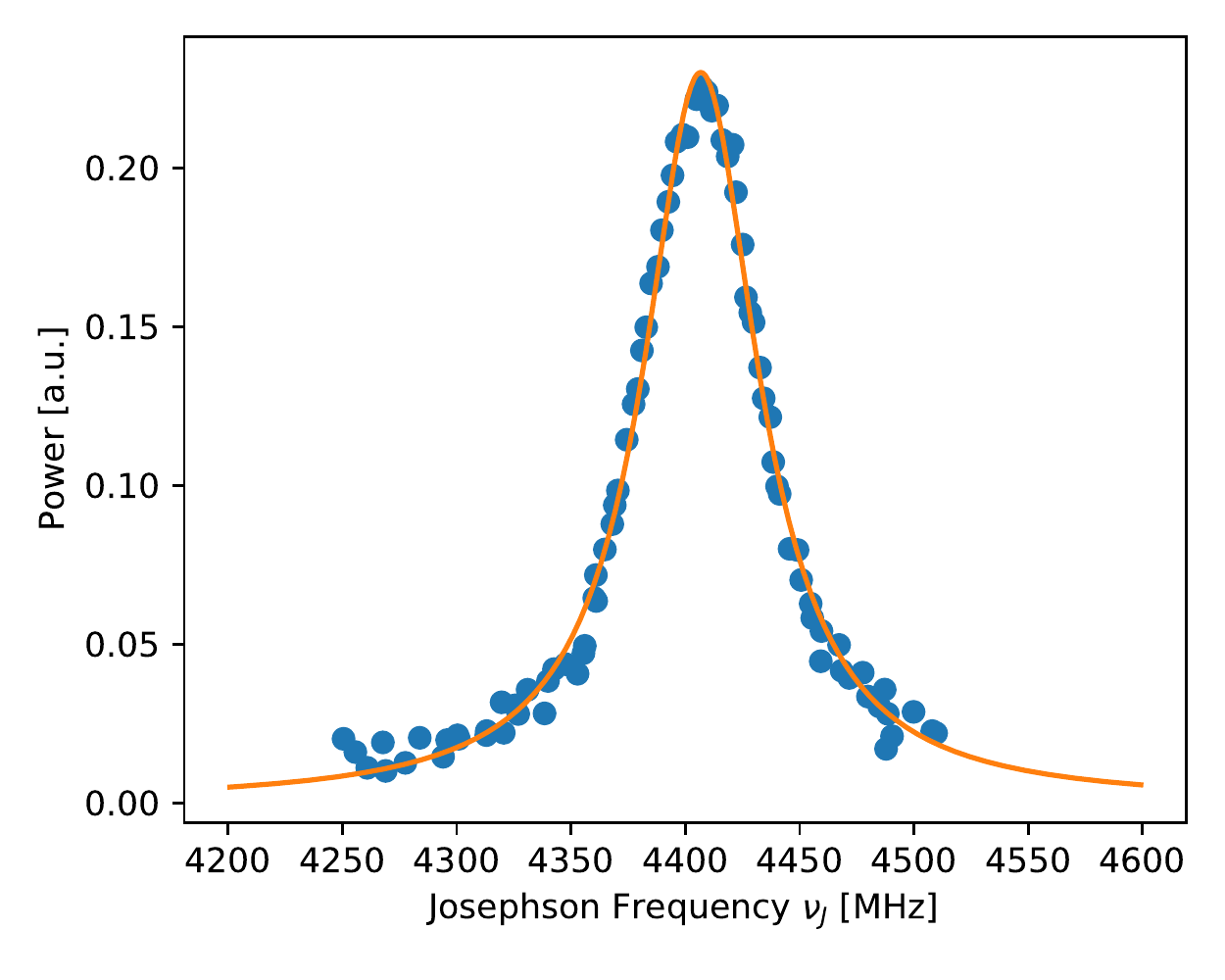}
    \caption{\textbf{Characterization of the resonator:} Measured total emitted power (dots) as a function of the central frequency of the emitted spectrum. The orange line is a Lorentzian fit yielding the resonator frequency $\nu_R= 4406.7~\pm~0.25$~MHZ and quality factor $Q=72\pm1.4$ (run 3).}
    \label{sup_resonator}
\end{center}
\end{figure}

Both runs 2 and 3 started with the determination of the resonator parameters. This was done by recording the PSD for many values of the bias voltage $V$ around $V_{k=1}$ (that is many values of the Josephson frequency $\nu_J$). The total emitted power was then computed as indicated in the previous appendix. This power being proportional to the real part of the resonator impedance $\mathrm{Re}[Z(\nu_J)]$, plotting it as a function of the central frequency $\nu_J$ of each spectrum reconstructs the resonator line shown in Fig. \ref{sup_resonator} (for run 3). Fitting the resonator line by a Lorentzian peak yields the central frequency of the resonator $\nu_R = 4406.75\pm 0.25$~MHz and a FWHM of $61.05\pm1.2$~MHz corresponding to a quality factor $Q=72\pm1.4$ (run 3). The same procedure yields $Q=36.6\pm0.7$ for run 2.

\section*{Appendix D: Estimation of the maximum Josephson energy (run 3)}
\begin{figure}[h!]
    \centering
    \includegraphics[width = 0.5\textwidth]{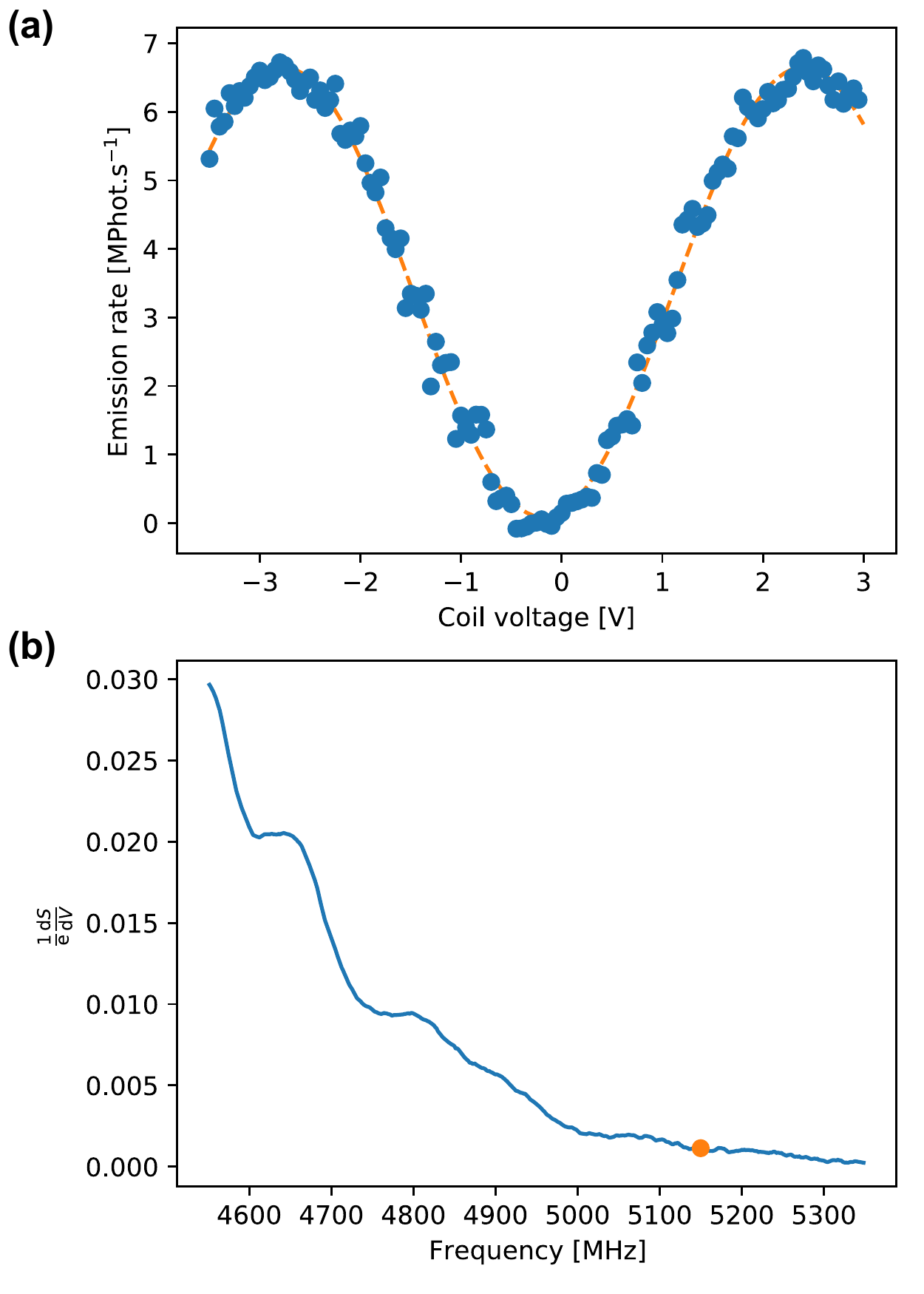}
    \caption{\textbf{Calibration of $E_J$ via the field dependence of the emission:} (a) Photon emission rate $\Gamma$ measured (dots) as a function of the voltage applied to the coil, recorded at $\nu_J = 5.15\pm 0.01$~GHz, far from the resonance frequency of the resonator. The dashed orange line is a sinusoidal fit. (b) Derivative of the shot-noise signal as a function of the frequency. The orange dot shows the point at $\nu_J$ used for dividing $\Gamma$ (see text).}
    \label{sup_ej_calib}
\end{figure}

In order to estimate the absolute maximal Josephson energy of our SQUID, we measure the emission at k=1 at a bias voltage corresponding to $\nu_\mathrm{J} = 5.15\pm 0.01$~GHz (on the high-frequency tail of the resonance), in order to maintain a low brightness and have a total emission rate  given by
\begin{equation}
    \Gamma = \frac{2\pi^2( E_J^*)^2}{\hbar^2\nu_J}\frac{\mathrm{Re}[Z(\nu_J)]}{R_\mathrm{K}},
    \label{gamma}
\end{equation}
where $R_\mathrm{K}=h/\mathrm{e}^2$ is the resistance quantum and $E_\mathrm{J}^*$ the effective Josephson energy of the SQUID renormalized by the phase fluctuations of its environment. We record $\Gamma$ as a function of the magnetic field [see Figure \ref{sup_ej_calib}.(b)] by sweeping the coil voltage in a single direction, extremely slowly and with very small steps, in order to avoid the lag and hysteresis mentioned in the main text and documented in the next section. As $E_J^*$ varies as the absolute value of a cosine function of the magnetic field, $\Gamma$ varies sinusoidally as expected, and a sinusoidal fit yields a precise value of the maximum emission rate $\Gamma_\mathrm{max}$. To eliminate the unknown $\mathrm{Re}[Z(\nu_J)]$ in Eq. \ref{gamma} and obtain $E_{J,max}^*$, we also measure the power spectral density $S(\nu)$ of the shot noise emitted when biasing the circuit at a voltage $V$ well above twice the superconducting gap voltage ($\sim$ 200 µV). We then divide $\Gamma_{max}$ by the derivative
\begin{equation}
    \frac{dS(\nu)}{dV} =2\mathrm{e}\frac{\mathrm{Re}[Z(\nu)]}{R_\mathrm{N}},
\end{equation}
at frequency $\nu_J$ [see Fig. \ref{sup_ej_calib}.(a)], with $R_\mathrm{N}$ the normal resistance of the SQUID that we estimate to be $335\pm6$~k$\Omega$ in run 3. The Josephson energy is thus simply given by
\begin{equation*}
    E_\mathrm{J}^{*2} = \frac{\Gamma_\mathrm{max}}{dS/dV}\frac{\mathrm{e}h^2\nu}{\pi^2}\frac{R_\mathrm{K}}{R_\mathrm{N}}.
\end{equation*}

We find $E_{J,\mathrm{max}}^*=1.01\pm0.02~\mu$eV, to be compared to the Ambegaokar-Baratoff value $E_{J,\mathrm{max}}^\mathrm{AB} = 1.73\pm 3~\mu$eV, which indicates a renormalization factor of 0.583 in perfect agreement with our estimate of 0.587 (see supplementary material of \cite{PhysRevLett.122.186804} for the method).

\section*{Appendix E: Hysteretic magnetic behavior of the SQUID and fitted $E_J$ values (run 2)}

As is the case for many Josephson devices, our sample suffers from the presence of magnetic vortices trapped in the superconducting electrodes in the vicinity of the SQUID. These vortices add a contribution to the external magnetic field applied by the coil. As the coil field is ramped the vortices can move and induce jumping or lagging of the effective flux experienced by the SQUID loop. Because the effective Josephson energy depends on this effective flux, its determination is problematic.

\begin{figure}[h!]
\begin{center}
    \includegraphics[width = 0.47\textwidth]{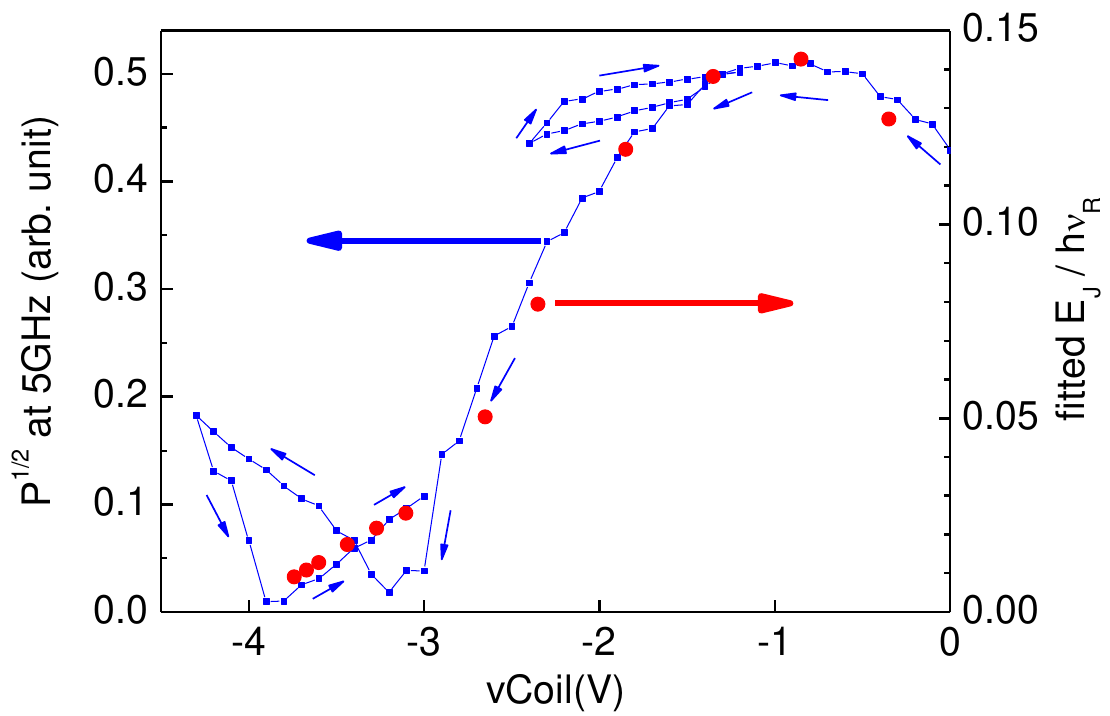}
    \caption{\textbf{Magnetic hysteresis and fitted $E_j$ values:} Square root of the emitted power $\sqrt{P}$ at 5~GHz (blue points, left axis) versus applied coil voltage  $v_\mathrm{Coil}$, and fitted $E_J$ values (red, right axis) presented in Fig. 2 of the main text. A relative vertical scaling is applied to compare the two datasets.}
    \label{hysteresis}
\end{center}
\end{figure}

However, as explained in the previous section, because the microwave emission is simply proportional to the square of the Josephson energy $E_J$ in the low occupation limit, $E_J$ variations can be followed from the measured emitted power when the bias voltage (or equivalently the Josephson frequency $\nu_J$ ) is tuned far on the tail of the resonator resonance. Figure \ref{hysteresis} shows a record of the square root of the emitted power $\sqrt{P}$ for $\nu_J=5~\rm{GHz}$, when sweeping back and forth the magnetic field along a particular path indicated by the blue arrows: starting from  of $V=0$~V, the voltage $v_\mathrm{Coil}$ applied to the coil circuit goes down to -2.5~V, increases to -1.5~V, goes all the way down to -4.2 V and finally re-increases to -3~V: a hysteretic emission is observed with lagging in both directions. To check that the $E_j$ values fitted in run 2 make sense and are valid, we plot them as a function of $v_\mathrm{Coil}$ and apply a relative vertical scaling to compare them to $\sqrt{P}$. The fitted $E_J$ values fall on the recorded trajectory for the emission, which shows that they are consistent with the chosen $v_\mathrm{Coil}$ values.

\section*{Appendix F: Bias voltage noise}
\label{section_noise}

The fridge setups with which we performed our experiments presented a bias voltage noise with a standard deviation of about 4~nV in run 1 and 2, and 80~nV in run 3. This transcribes into a Josephson frequency noise and a finite emission width (at $k=1$) of less than 2 MHz and about 38 MHz, respectively, to be compared to the resonator linewidths of about 120~MHz and 60~MHz, respectively. Consequently, we could neglect this noise in our analysis for run 2, but had to take it into account for run 3. Indeed, when the circuit is biased at $V_0 =h\nu_{J0}/2\mathrm{e}$, the power spectral density at a given instant and at a frequency $\nu$ around $\nu_{J0}$ is proportional to the product of the impedance of the resonator at that frequency $\nu$ by the probability to produce a photon at frequency $\nu = \nu_{J0}+2\mathrm{e}\delta V/h$:
\begin{equation*}
    \mathrm{PSD (\nu)}\propto \mathrm{Re}[Z(\nu)]\times P(h\nu_{J0}+2\mathrm{e}\delta V),
\end{equation*}
with $\delta V$ the noise voltage at that instant.

\begin{figure}[h!]
\begin{center}
    \includegraphics[width = 0.45\textwidth]{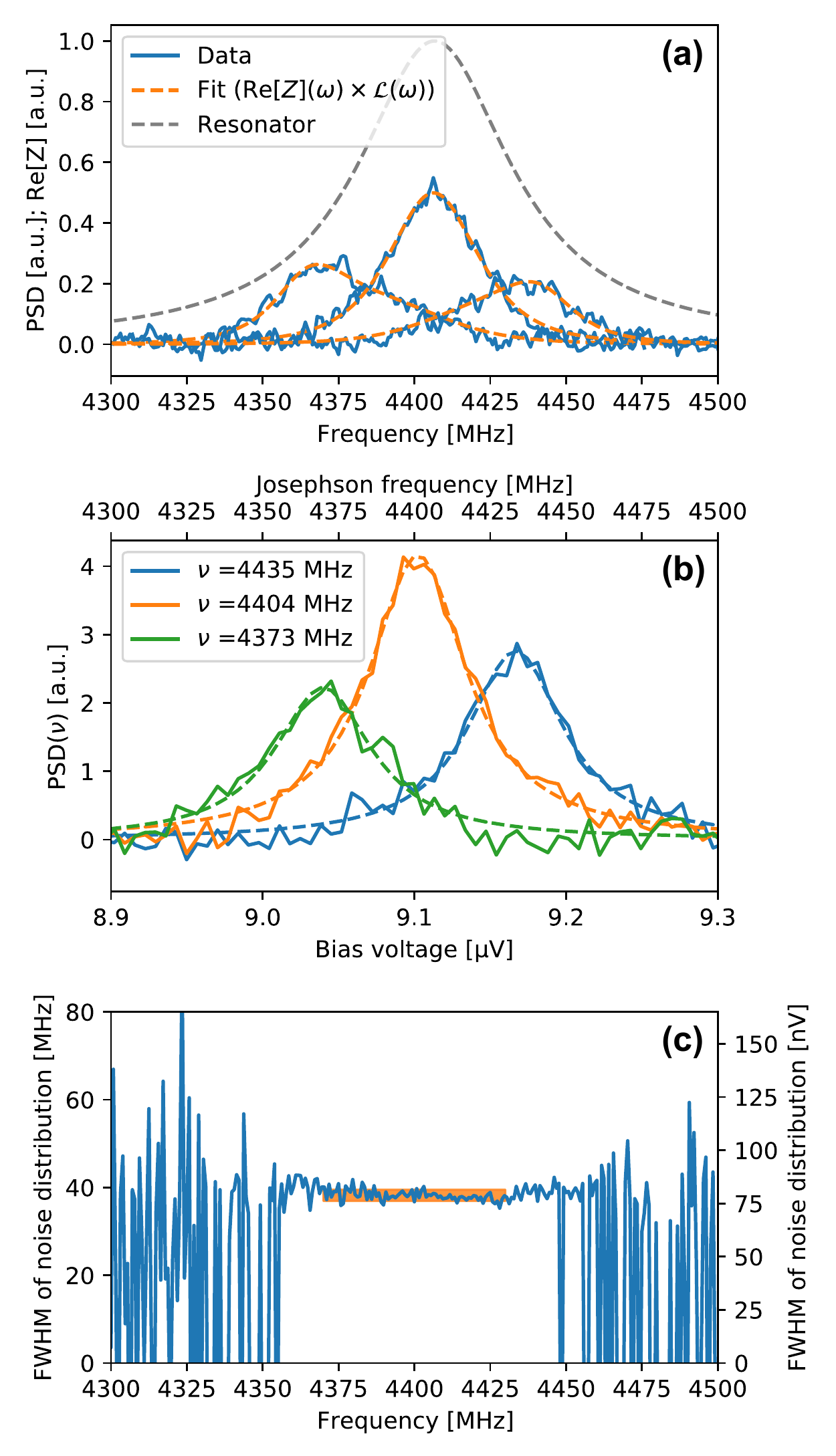}
    \caption{\textbf{Fitting of the voltage noise:} (a) Example of three measured PSD of emission (blue lines) taken at various voltage biases, below, at, and above the resonator central frequency, as well as their fit (orange) by a product of two Lorentzian lines (see text). The resonator line shape (dashed grey line) is superposed to allow the reader to better locate the bias values. (b) Three examples of emitted power P (solid lines) taken at fixed frequencies $\nu$ as a function of the bias voltage $V$, as well as their Lorentzian fit (dashed lines). (c) FWHM extracted from many $P(V)$ curves as those shown in panel b, on a dense bias grid. The orange bar shows the window used to compute the mean value of the emission width and its height indicates the corresponding standard deviation.
    }
    \label{sup_noise}
\end{center}
\end{figure}

In the case of a purely thermal noise with Gaussian statistics, one would assume the PSD at $k=1$ to be the product of the Lorentzian shape of the resonator determined in Appendix C, centered on $\nu_R$, by a Gaussian centered on $\nu_{J0}$. However, fitting the measured PSD revealed that also using a Lorentzian shape $\mathcal{L}$ for the noise distribution yields better results [see fig. \ref{sup_noise}.(a)].

Now considering the emission at fixed frequency $\nu$ as a function of the bias voltage V (or $\nu_J$), allows us to measure more directly the noise distribution [see the three examples of Fig. \ref{sup_noise}(b)], as the impedance $Z(\nu)$ becomes a simple multiplicative constant to a V-dependant line shape. Once again, a Lorentzian fit is in better agreement with the experimental data than a Gaussian. Figure \ref{sup_noise}(c) shows the FWHM of the noise distribution  (expressed in frequency units), measured in this way on many curves at different biases: its value over a window centered on the resonator is $38.2\pm 1.3 $~MHz.

\section*{Appendix G: Set of measured $g^{(2)}$ functions and Fano factors (run 3)}
\label{section_g2}
\begin{figure*}
\begin{center}
    \includegraphics[width = 0.8\textwidth]{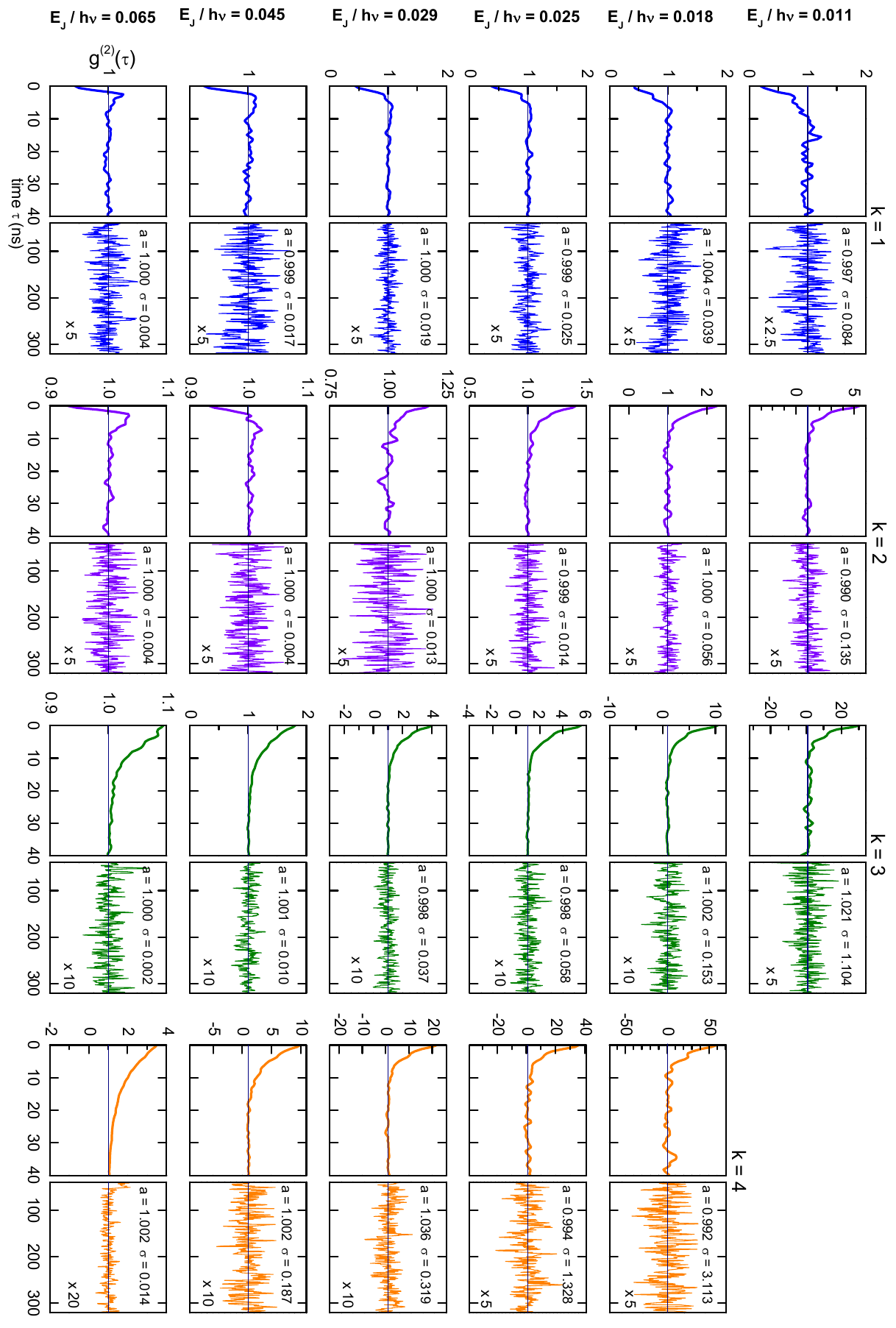}
    \caption{\textbf{Function $g^{(2)}(\tau)$ for various values of $k$ and $E_J$:} Raw experimental measurement of $g^{(2)}(\tau)$ from $k = 1$ to $k=4$ (from left to right) and for $E_J/h\nu_r = \{0.011,0.018,0.025,0.029,0.045,0.065\}$. The left panel of each panel pair presents the short time variation of $g^{2}$ while the right panel presents the noise on the measured $g^{(2)}(\tau)$ fluctuating around 1 at long time. A vertical multiplication factor is indicated in these right panels together with the average value $a$ and the standard deviation $\sigma$ of the noise.}
    \label{g2set}
\end{center}
\end{figure*}

Figure \ref{g2set} presents the full set of $g^{(2)}(\tau)$ data, shown both for short and long times. Antibunching of the $k=1$ case can be observed in the dip of $g^{(2)}(\tau)$ for $k=1$ at all $E_\mathrm{J}$ values, as well as in the data taken at strong $E_\mathrm{J}/h\nu = 0.045$ and $0.065$, for $k=2$. We extract $F(k, E_J)$ by integrating $g^{(2)}(\tau)$ according to 
\begin{equation}
 F_k = 1+2\Gamma_k\int_{0}^{+\infty}{\left[1-g^{(2)}(\tau)\right]\mathrm{d}\tau},
\end{equation}
over the 40 ns time window shown in the left panels of Fig. \ref{g2set}.
The relative Fano factor uncertainty is determined by the standard deviation $\sigma$ of $g^{(2)}$ measured at long time (right panels) and the total number $N$ of integrated points: $\Delta F_k/F_k= (1-1/F_k) (\Delta \Gamma_k/\Gamma_k +2\Gamma_k \sqrt{N} \sigma/F_k)$.

\section*{Appendix H: Including the effects of voltage noise in quantum numerical simulations \label{Sec:sup_noise}}

\begin{figure}[h!]
\begin{center}
    \includegraphics[width = 0.47\textwidth]{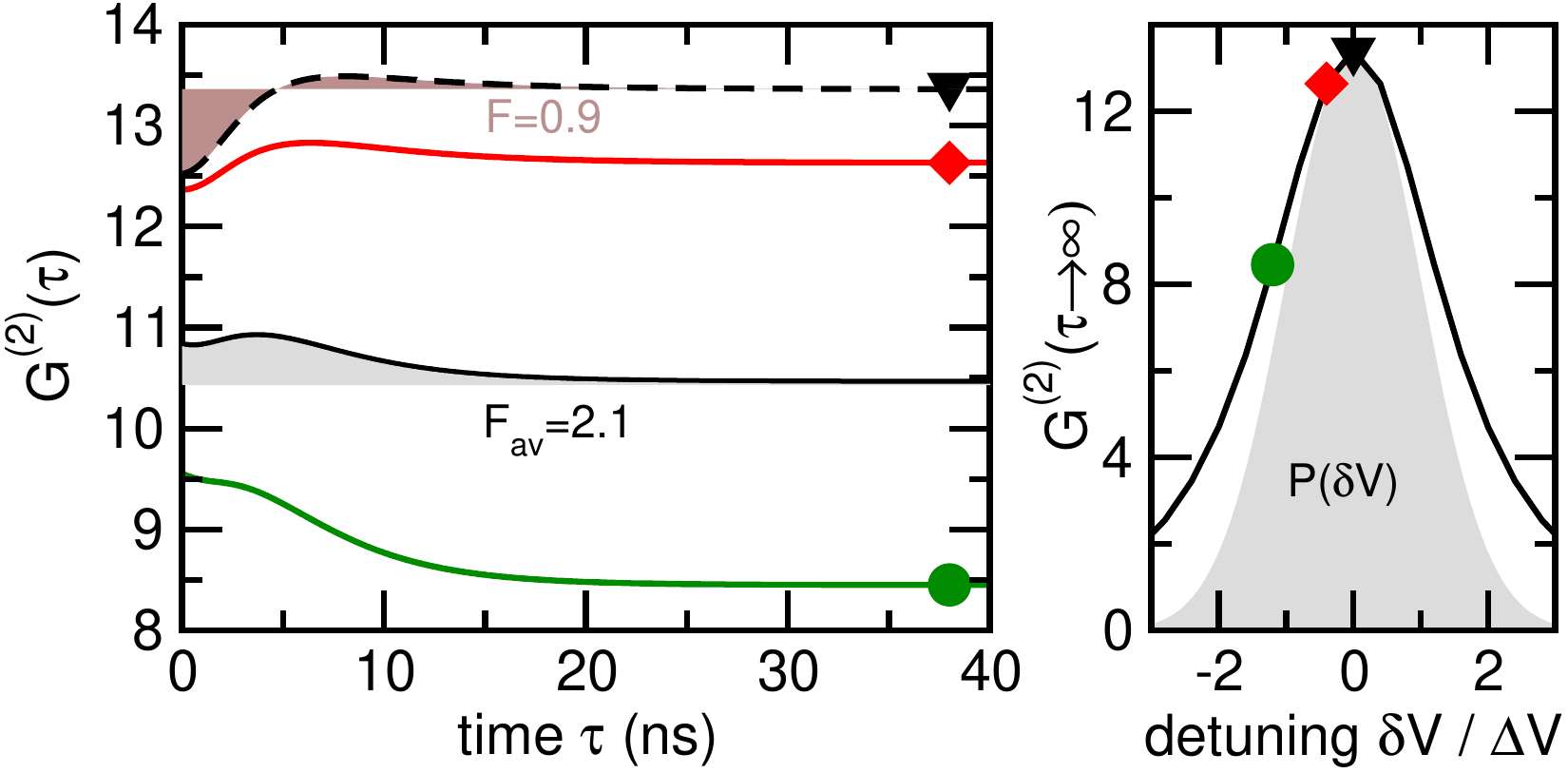}
    \caption{\textbf{Effects of voltage noise on the correlation function $G^{(2)}(\tau)$} are simulated by averaging over a Gaussian normal distribution $P(\delta V/\Delta V)$ (gray shaded in the right panel) with the variance extracted from experiment as described in Sec. \ref{section_noise} above. The unnormalized correlation functions $G^{(2)}(\tau)$ for three detunings  ($0,\, 0.2,$ and $0.4\Delta V$, see markers) are shown in the left panel, together with the averaged result (black). The areas under the curves (shaded in the left panel) enter the Fano factor according to Eq. (\ref{Bk}) of the main text (other parameters: $k=3$, $E_J= 0.9 E_J^\textrm{sat.}=0.07 /h\nu$).
        \label{sup_noise_G}
}\end{center}
\end{figure}

This section describes how fluctuations of the bias-voltage affect the dynamics of a Josephson-photonics system and how they can be accounted for in a simulation of measured observables. Obviously, all observables discussed here, from the resonator occupation and emitted power to correlation functions and Fano factors, depend on the detuning from the voltage matching condition,  $2\mathrm{e}V=k\hbar \omega$, of the $k$-photon resonance. Examples are the measured emitted power in Fig. 1(c) of the main text, or the simulated results in Fig.~\ref{sup_noise_G}, which show the unnormalized correlation function $G^{(2)}(\tau)=\left<n\right>^2\times g^{(2)}(\tau)$ with the dependence of $G^{(2)}(\tau \rightarrow \infty) = \langle n \rangle^2$ on the detuning illustrated in the right panel. In an experiment, even if the mean voltage is tuned to resonance, voltage fluctuations will let the system explore the effects of such detuning. Voltage fluctuations are always present and have been carefully measured and characterized for the various runs of our experiments (see Appendix F above). The typical size of  fluctuations has been determined and is used to scale the voltage in the right panel of Fig.~\ref{sup_noise_G}. In fact, that typical size (in run 3) is not small, so that there is a substantial effect on the measured observables for typical fluctuations $\Delta V$, as shown by the markers in Fig.~\ref{sup_noise_G} right indicating detunings of  $0,\, 0.2,$ and $0.4~\Delta V$. 

\begin{figure}[!ht]
\begin{center}
    \includegraphics[width = 0.47\textwidth]{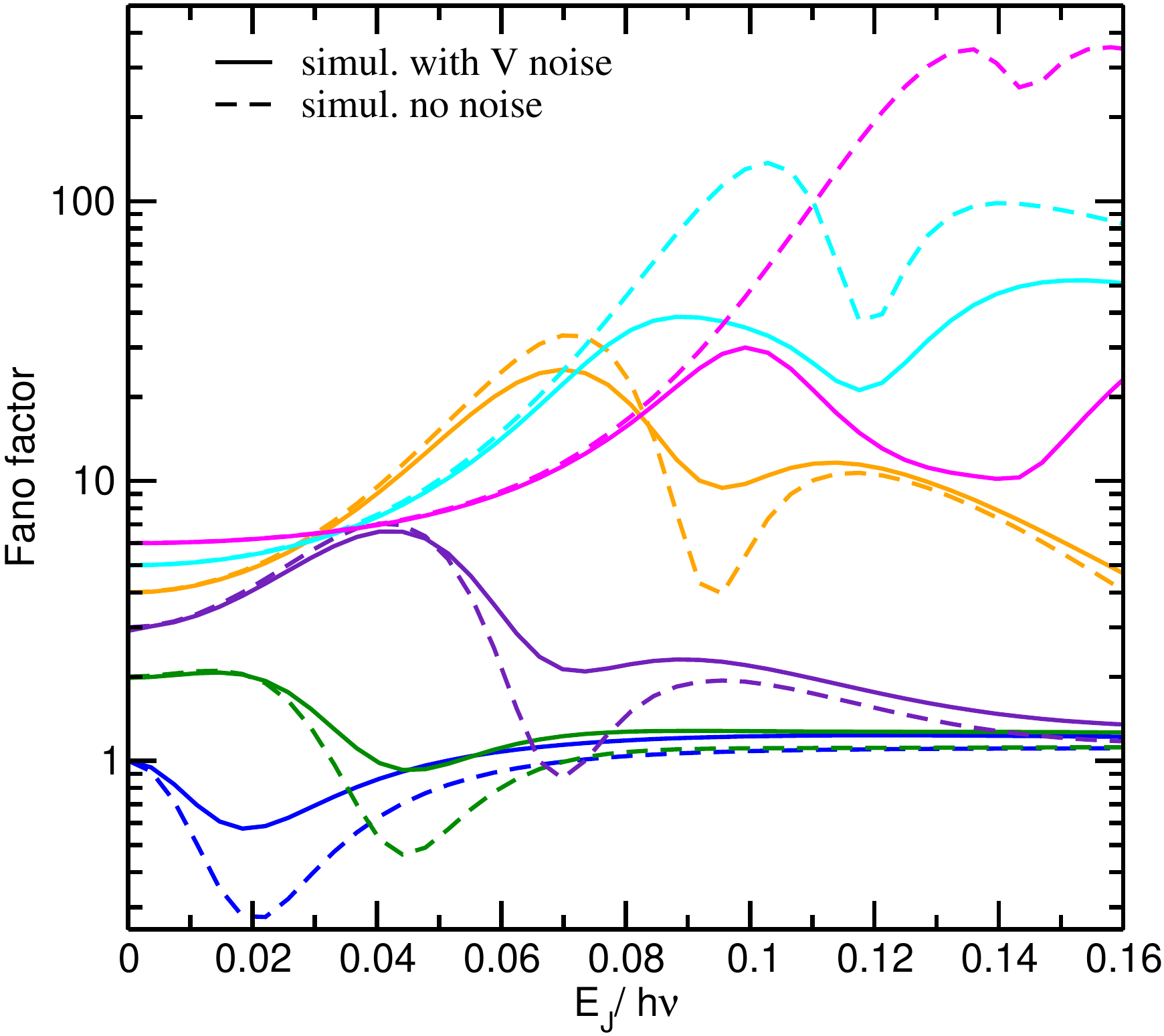}
       \caption{\textbf{Effects of voltage noise on the Fano factor:} Driving dependence of the Fano factor for the resonances $k=1,2,\hdots 6$ as can be read of by the weak driving limit simulated without noise (for zero nominal detuning) and including noise effects as described by Fig. (2)).
        \label{sup_noise_F}
}\end{center}
\end{figure}

To account for the main effects of voltage fluctuations, one can assume that they are of thermal origin and of classical nature. While fluctuations are much slower than the excitation and leaking dynamics of the resonator, the  averaging time over which measurements are assembled is much longer. In this quasi-static case, voltage fluctuations can be accounted for by averaging over a distribution of detunings. Assuming a classical noise model, we chose a Gaussian distribution. This noise model has been successfully used to prove and characterize entanglement of a two-resonator Josephson-photonics setup in \cite{Peugeot2021}, but different or refined methods (see e.g. the S.M. of ~\cite{Armour2017}) are also possible. 

We find that voltage noise has a substantial effect in our experiment. Already, on mean observables, such as the mean number of photons in the resonator $\langle n \rangle$ (cf. Fig. 3(a) of the main text), but also on correlation functions (left panel of Fig.~\ref{sup_noise_G}) and on the Fano factor, see Fig.~\ref{sup_noise_F}.
The left panel of Fig.~\ref{sup_noise_G} highlights a situation where that effect is very pronounced. As may be expected, this happens around the transition points of the semi-classical results discussed in Appendix I below (which, in fact, show non-analyticity only for the resonant case). Shown are the (unnormalized) correlation functions $G^{(2)}(\tau)$ for the three values of detunings indicated in the right panel, as well as the result of averaging $G^{(2)}(\tau)$ over the distribution $P(\delta V)$, shown shaded in the right panel. The averaging completely changes the time-dependence of $G^{(2)}(\tau)$, and can even result in a super-Poissonian Fano factor, $F_\textrm{av}>1$, where the noiseless resonant Fano factor would be sub-Poissonian, $F<1$, represented by the shaded (positive and negative) areas and cf. Fig.~\ref{sup_noise_F}.

In contrast to the strong effects on the dip-peak structure at larger driving, for weak driving (and the naive $F=k$ results) the effect of voltage fluctuations is nearly negligible. In particular, this can be observed (see Fig.~\ref{sup_noise_F}) for higher resonances, where the frequency mismatch fluctuations $\propto \Delta V/k$ are reduced. Altogether, classical fluctuations, not unlike the quantum fluctuations due to the large $\alpha$, lead to a broadening of the pronounced features predicted by semi-classics, but the distortion of features can be quite strong. Despite the large impact of fluctuations, the highly nontrivial, complex driving dependence of the experimental and simulated Fano factors matches astonishingly well. Figure 3(b) of the main text demonstrates the high degree of theoretical understanding and experimental control Josephson-photonics systems offer in exploring this novel regime of strong-coupling quantum electrodynamics.

\section*{Appendix I: Strong emission regime of a Josephson-photonics system at a multi-photon resonance \label{Sec:sup_strong_driving}}
The most striking feature of biasing a Josephson-photonics system at a multi-photon resonance is the emission of photon multiplets. This bunching is reflected in a photon Fano factor $F_k=k$, observed for the $k$-photon resonance at weak driving in accordance with the naive expectation. However, as Josephson energy and therefore emission get stronger and the resonator is not relaxing to its ground state between consecutive Cooper-pair tunneling events, the dynamics becomes more complex and the behavior of the Fano factor of photonic emission [Fig. 3(c) of the main text] and the mean resonator occupation [Fig. 2(b) of the main text] become highly nontrivial. In essence, this is a consequence of two competing effects in the nonlinear terms of the Hamiltonian: (i) at moderate $E_J$ the $k$-parametric drive term, $\sim a^k + (a^\dagger)^k$ in Eq. (\ref{H_rwa}), is strongly superlinear;  in the fashion of stimulated emission, the tunneling matrix elements and the corresponding excitation rate are strongly enhanced, if there are already excitations present in the resonator. (ii) At stronger $E_J$, however, the Josephson nonlinearity formally encoded in the $B_k$ operators in Eq. (3) suppresses the efficiency of the driving. In the classical limit, this is reflected in a Bessel function reaching its maximum \cite{armour2013universal}, while for large $\alpha$ the same nonlinearity appears on a few-photon level and the resonator is effectively reduced to a few-level system (cf. Ref.~\cite{PhysRevLett.122.186804} for the case $k=1$).

\begin{figure}[!ht]
\begin{center}
    \includegraphics[width = 0.48\textwidth]{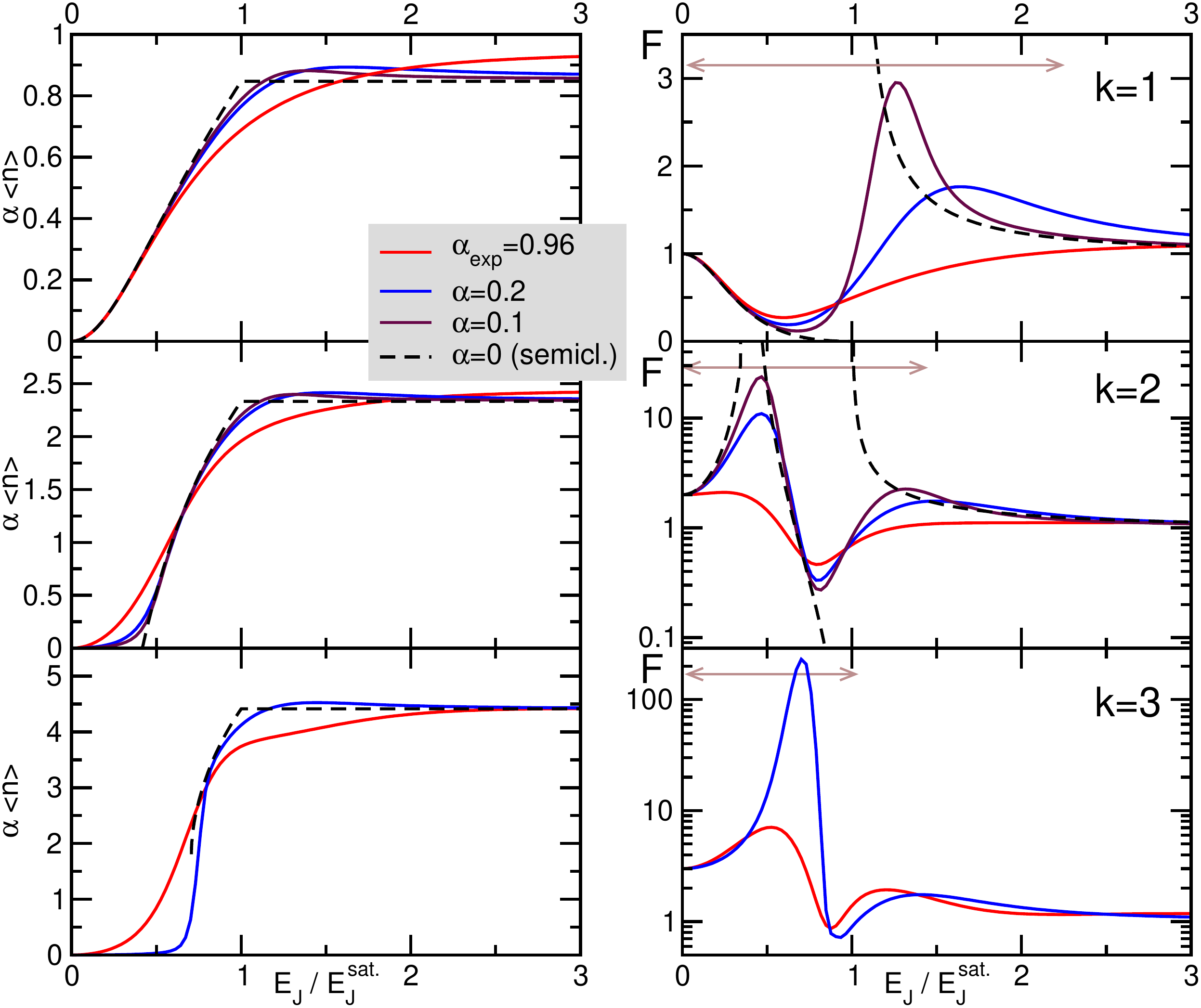}
    \caption{\textbf{Strong emission at multi-photon resonances, $k=1,\,2,\,3$:} Scaled occupation $\alpha \langle n\rangle$ (left) and Fano factor F of photon emission (right). Beyond the weak driving limit, the nonlinear dynamics is complex and shows two transitions (for $k\ge2$): a saturation transition at $E_J/E_J^\textrm{sat.}\equiv 1$ and a parametric transition at lower $E_J$ (see Ref.~\cite{armour2013universal} for derivations and details). While transitions are non-analytic in the semi-classical limit, $\alpha \rightarrow 0$ (dashed), higher values of $\alpha$ progressively smoothen them and yield emission below the parametric threshold according to a rate picture of Dynamical Coulomb blockade [cf. Eq.~(3) and Fig. 2(b) of the main text]. The Fano factor starting form the naive bunch-size, $F=k$, goes through a peak around the parametric threshold and a dip-peak structure around the saturation transition to settle at $F=1$ for very strong driving. The ranges of drivings shown in Fig. 3(c) of the main text are marked by arrows.}
    \label{sup_strong_driving_overview}
\end{center}
\end{figure}

Understanding the system in the semi-classical limit, $\alpha \rightarrow 0$  offers some insight into the complex dynamics, even for our case, where $\alpha \sim 1$. The generic dependence of the resonator occupation and of the Fano factor on $E_J$ is visualized in Fig. \ref{sup_strong_driving_overview}. A detailed semi-classical analysis \cite{armour2013universal} shows that the scaled resonator occupation $\alpha \langle n \rangle$ undergoes a (bifurcation) transition to a saturation value for all $k$ corresponding to the nonlinear suppression discussed as (ii) above. The corresponding threshold value $E_{J}^{sat.}$ above which $\alpha \langle n\rangle$ saturates, has been discussed and derived in detail in \cite{armour2013universal}. For smaller $E_J$, the first and second resonance, $k=1$ and $k=2$, differ from $k \ge 3$. 

\emph{For the $k=1$ resonance}, the occupation continuously increases with the $E_J$ (quadratically, as expected in the weak $E_J$ limit, where the Josephson coupling reduces to a linear drive). 
\emph{For the conventional parametric ($k=2$) resonance}, emission is suppressed for low $E_J$ until a continuous onset above a parametric threshold value \cite{armour2013universal}, while \emph{for $k\ge 3$} at threshold the occupation jumps from zero to a finite value in the semi-classical limit. In Fig. 2(b) of the main text, which (because $\alpha \sim 1$) is far from the semi-classical limit, the parametric behavior manifests as a clear upward turn of the numerical results (solid) compared to the rate equation result of Eq.~(3) in the main text (dashed), valid only in the low $E_J$ limit. 

For the Fano factor, one generically finds a peak around the parametric threshold, which has been described for $k=2$ as bursts \cite{PhysRevB.86.054514}, each encompassing multiple pairs of photons. A semi-classical analysis is possible beyond the parametric threshold and finds a sudden switching of the noise. Noise and Fano factor vanish just below the saturation threshold, while they diverge above. This behavior has been explained in \cite{Armour2017} as a generic result of a certain type of nonlinear driving Hamiltonian, where the nature of the fixed point and the corresponding fluctuation properties (amplitudes and correlation times) change abruptly.

The impact of quantum fluctuations for large $\alpha$ masks all sharp transitions predicted by semiclassics. Besides blurring all transitions, quantum fluctuations may also allow dynamical transitions between solutions and crucially affect certain observables \cite{LangNJP2021}. 

\section*{Appendix J: Simulated intra-resonator Wigner functions} \label{Sec:Wigner}

\begin{figure}[!ht]
\begin{center}
    \includegraphics[width = 0.49\textwidth]{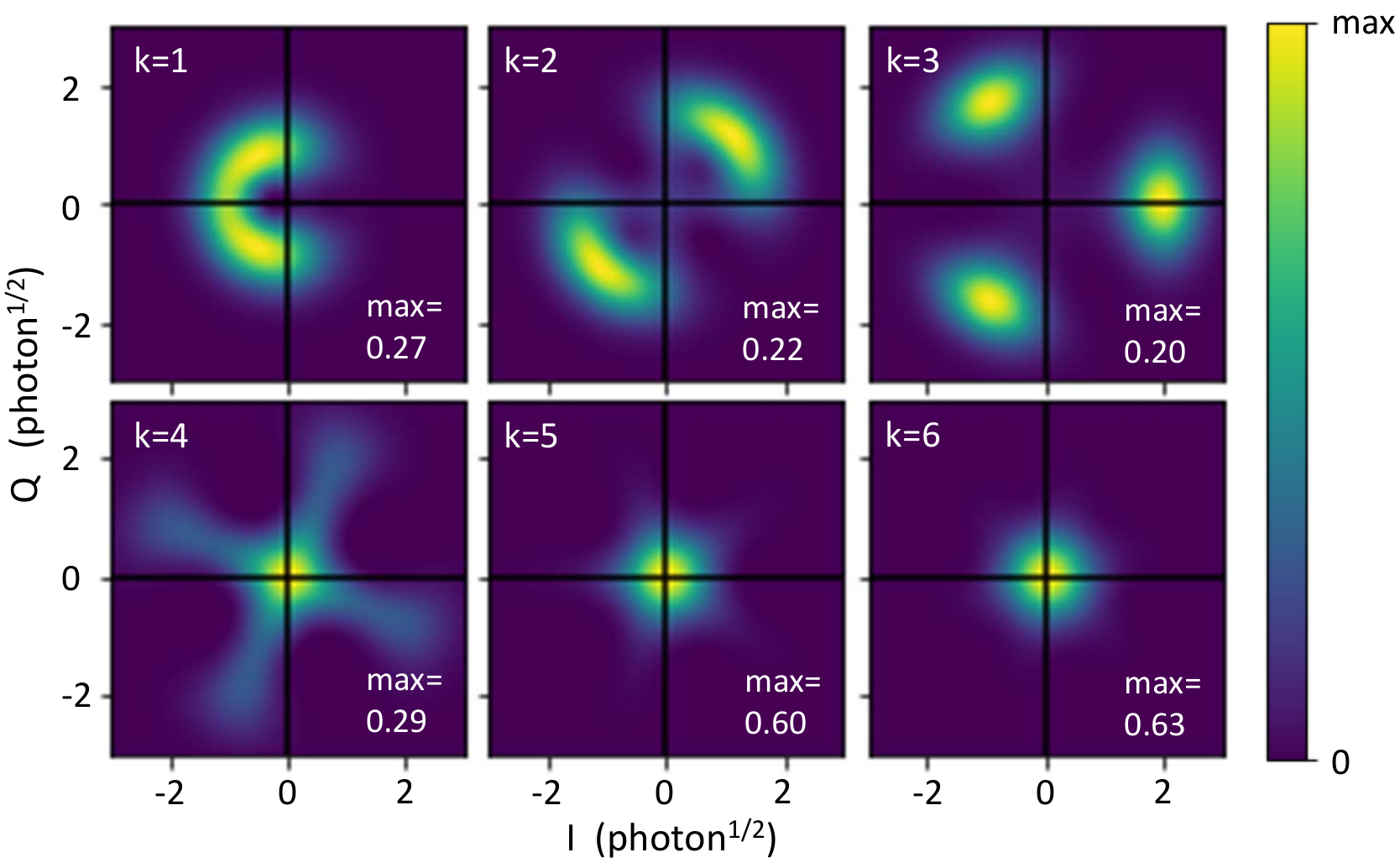}
    \caption{\textbf{Simulated intra-resonator field Wigner function} with the parameters of run 2, at $E_J/h \nu_R = 0.142$: amplitude of the Wigner function as a function of the two reduced quadratures of the field, for $k=$1 to 6. Note that the colorscale is bounded by the minimal value 0 indicating the absence of predicted Wigner negativity. The text in each panel indicates the maximal value of the color scale.}
    \label{Wigner}
\end{center}
\end{figure}

The k-granularity of the microwave emission is not the only quantum feature of the emission process described in this work. Although we did not measure it, the field statistics is also non Gaussian.  The intra-resonator Wigner function has a  k-fold symmetry, but apparently no Wigner negativity. Figure ~\ref{Wigner} displays for instance the Wigner functions obtained from the master equation simulations done for Fig. \ref{fig2}(b) at the highest Josephson energy $E_J/h \nu_R = 0.142$ [vertical dotted line in the figure]. The k symmetry is clearly visible from k=2 to 5, the very low occupation at k=6 making the 6-fold symmetry barely visible. At $k=1$, the system is close to saturation and the Wigner function gets deformed compared to the displaced ground state  simulated at low $E_J$ (not shown), as for a linearly driven damped oscillator.

\bibliography{biblio}

\begin{thebibliography}{41}%
\makeatletter
\providecommand \@ifxundefined [1]{%
 \@ifx{#1\undefined}
}%
\providecommand \@ifnum [1]{%
 \ifnum #1\expandafter \@firstoftwo
 \else \expandafter \@secondoftwo
 \fi
}%
\providecommand \@ifx [1]{%
 \ifx #1\expandafter \@firstoftwo
 \else \expandafter \@secondoftwo
 \fi
}%
\providecommand \natexlab [1]{#1}%
\providecommand \enquote  [1]{``#1''}%
\providecommand \bibnamefont  [1]{#1}%
\providecommand \bibfnamefont [1]{#1}%
\providecommand \citenamefont [1]{#1}%
\providecommand \href@noop [0]{\@secondoftwo}%
\providecommand \href [0]{\begingroup \@sanitize@url \@href}%
\providecommand \@href[1]{\@@startlink{#1}\@@href}%
\providecommand \@@href[1]{\endgroup#1\@@endlink}%
\providecommand \@sanitize@url [0]{\catcode `\\12\catcode `\$12\catcode
  `\&12\catcode `\#12\catcode `\^12\catcode `\_12\catcode `\%12\relax}%
\providecommand \@@startlink[1]{}%
\providecommand \@@endlink[0]{}%
\providecommand \url  [0]{\begingroup\@sanitize@url \@url }%
\providecommand \@url [1]{\endgroup\@href {#1}{\urlprefix }}%
\providecommand \urlprefix  [0]{URL }%
\providecommand \Eprint [0]{\href }%
\providecommand \doibase [0]{http://dx.doi.org/}%
\providecommand \selectlanguage [0]{\@gobble}%
\providecommand \bibinfo  [0]{\@secondoftwo}%
\providecommand \bibfield  [0]{\@secondoftwo}%
\providecommand \translation [1]{[#1]}%
\providecommand \BibitemOpen [0]{}%
\providecommand \bibitemStop [0]{}%
\providecommand \bibitemNoStop [0]{.\EOS\space}%
\providecommand \EOS [0]{\spacefactor3000\relax}%
\providecommand \BibitemShut  [1]{\csname bibitem#1\endcsname}%
\let\auto@bib@innerbib\@empty
\bibitem [{\citenamefont {Cohen-Tannoudji}\ \emph {et~al.}(1989)\citenamefont
  {Cohen-Tannoudji}, \citenamefont {Dupont-Roc},\ and\ \citenamefont
  {Grynberg}}]{alpha}%
  \BibitemOpen
  \bibfield  {author} {\bibinfo {author} {\bibfnamefont {C.}~\bibnamefont
  {Cohen-Tannoudji}}, \bibinfo {author} {\bibfnamefont {J.}~\bibnamefont
  {Dupont-Roc}}, \ and\ \bibinfo {author} {\bibfnamefont {G.W.}\ \bibnamefont
  {Grynberg}},\ }\href@noop {} {\emph {\bibinfo {title} {Introduction to
  Quantum Electrodynamics}}}\ (\bibinfo  {publisher} {Wiley-Interscience},\
  \bibinfo {year} {1989})\BibitemShut {NoStop}%
\bibitem [{\citenamefont {Corona}\ \emph {et~al.}(2011)\citenamefont {Corona},
  \citenamefont {Garay-Palmett},\ and\ \citenamefont
  {B.}}]{PhysRevA.84.033823}%
  \BibitemOpen
  \bibfield  {author} {\bibinfo {author} {\bibfnamefont {M.}~\bibnamefont
  {Corona}}, \bibinfo {author} {\bibfnamefont {K.}~\bibnamefont
  {Garay-Palmett}}, \ and\ \bibinfo {author} {\bibfnamefont {U’Ren~A.}\
  \bibnamefont {B.}},\ }\bibfield  {title} {\enquote {\bibinfo {title}
  {Third-order spontaneous parametric down-conversion in thin optical fibers as
  a photon-triplet source},}\ }\href {\doibase 10.1103/PhysRevA.84.033823}
  {\bibfield  {journal} {\bibinfo  {journal} {Phys. Rev. A}\ }\textbf {\bibinfo
  {volume} {84}},\ \bibinfo {pages} {033823} (\bibinfo {year}
  {2011})}\BibitemShut {NoStop}%
\bibitem [{\citenamefont {Chang}\ \emph {et~al.}(2020)\citenamefont {Chang},
  \citenamefont {Sab{\'\i}n}, \citenamefont {Forn-D{\'\i}az}, \citenamefont
  {Quijandr{\'\i}a}, \citenamefont {Vadiraj}, \citenamefont {Nsanzineza},
  \citenamefont {Johansson},\ and\ \citenamefont
  {Wilson}}]{chang2020observation}%
  \BibitemOpen
  \bibfield  {author} {\bibinfo {author} {\bibfnamefont {CW~Sandbo}\
  \bibnamefont {Chang}}, \bibinfo {author} {\bibfnamefont {Carlos}\
  \bibnamefont {Sab{\'\i}n}}, \bibinfo {author} {\bibfnamefont {P}~\bibnamefont
  {Forn-D{\'\i}az}}, \bibinfo {author} {\bibfnamefont {Fernando}\ \bibnamefont
  {Quijandr{\'\i}a}}, \bibinfo {author} {\bibfnamefont {AM}~\bibnamefont
  {Vadiraj}}, \bibinfo {author} {\bibfnamefont {I}~\bibnamefont {Nsanzineza}},
  \bibinfo {author} {\bibfnamefont {G{\"o}ran}\ \bibnamefont {Johansson}}, \
  and\ \bibinfo {author} {\bibfnamefont {CM}~\bibnamefont {Wilson}},\
  }\bibfield  {title} {\enquote {\bibinfo {title} {Observation of three-photon
  spontaneous parametric down-conversion in a superconducting parametric
  cavity},}\ }\href@noop {} {\bibfield  {journal} {\bibinfo  {journal}
  {Physical Review X}\ }\textbf {\bibinfo {volume} {10}},\ \bibinfo {pages}
  {011011} (\bibinfo {year} {2020})}\BibitemShut {NoStop}%
\bibitem [{\citenamefont {Ingold}\ and\ \citenamefont
  {Nazarov}(1992)}]{ingold92}%
  \BibitemOpen
  \bibfield  {author} {\bibinfo {author} {\bibfnamefont {G.-L.}\ \bibnamefont
  {Ingold}}\ and\ \bibinfo {author} {\bibfnamefont {Y.~V.}\ \bibnamefont
  {Nazarov}},\ }\bibfield  {title} {\enquote {\bibinfo {title} {Charge
  tunneling rates in ultrasmall junctions},}\ }in\ \href@noop {} {\emph
  {\bibinfo {booktitle} {Single charge tunneling}}},\ \bibinfo {editor} {edited
  by\ \bibinfo {editor} {\bibfnamefont {H.}~\bibnamefont {Grabert}}\ and\
  \bibinfo {editor} {\bibfnamefont {M.~H.}\ \bibnamefont {Devoret}}}\ (\bibinfo
   {publisher} {Plenum},\ \bibinfo {year} {1992})\BibitemShut {NoStop}%
\bibitem [{\citenamefont {Lesovik}\ and\ \citenamefont
  {Loosen}(1997)}]{0803.0020}%
  \BibitemOpen
  \bibfield  {author} {\bibinfo {author} {\bibfnamefont {GB}~\bibnamefont
  {Lesovik}}\ and\ \bibinfo {author} {\bibfnamefont {R}~\bibnamefont
  {Loosen}},\ }\bibfield  {title} {\enquote {\bibinfo {title} {On the detection
  of finite-frequency current fluctuations},}\ }\href@noop {} {\bibfield
  {journal} {\bibinfo  {journal} {Journal of Experimental and Theoretical
  Physics Letters}\ }\textbf {\bibinfo {volume} {65}},\ \bibinfo {pages}
  {295--299} (\bibinfo {year} {1997})}\BibitemShut {NoStop}%
\bibitem [{\citenamefont {Grimsmo}\ \emph {et~al.}(2016)\citenamefont
  {Grimsmo}, \citenamefont {Qassemi}, \citenamefont {Reulet},\ and\
  \citenamefont {Blais}}]{PhysRevLett.116.043602}%
  \BibitemOpen
  \bibfield  {author} {\bibinfo {author} {\bibfnamefont {Arne~L.}\ \bibnamefont
  {Grimsmo}}, \bibinfo {author} {\bibfnamefont {Farzad}\ \bibnamefont
  {Qassemi}}, \bibinfo {author} {\bibfnamefont {Bertrand}\ \bibnamefont
  {Reulet}}, \ and\ \bibinfo {author} {\bibfnamefont {Alexandre}\ \bibnamefont
  {Blais}},\ }\bibfield  {title} {\enquote {\bibinfo {title} {Quantum optics
  theory of electronic noise in coherent conductors},}\ }\href {\doibase
  10.1103/PhysRevLett.116.043602} {\bibfield  {journal} {\bibinfo  {journal}
  {Phys. Rev. Lett.}\ }\textbf {\bibinfo {volume} {116}},\ \bibinfo {pages}
  {043602} (\bibinfo {year} {2016})}\BibitemShut {NoStop}%
\bibitem [{\citenamefont {Hofheinz}\ \emph {et~al.}(2011)\citenamefont
  {Hofheinz}, \citenamefont {Portier}, \citenamefont {Baudouin}, \citenamefont
  {Joyez}, \citenamefont {Vion}, \citenamefont {Bertet}, \citenamefont
  {Roche},\ and\ \citenamefont {Esteve}}]{PhysRevLett.106.217005}%
  \BibitemOpen
  \bibfield  {author} {\bibinfo {author} {\bibfnamefont {M.}~\bibnamefont
  {Hofheinz}}, \bibinfo {author} {\bibfnamefont {F.}~\bibnamefont {Portier}},
  \bibinfo {author} {\bibfnamefont {Q.}~\bibnamefont {Baudouin}}, \bibinfo
  {author} {\bibfnamefont {P.}~\bibnamefont {Joyez}}, \bibinfo {author}
  {\bibfnamefont {D.}~\bibnamefont {Vion}}, \bibinfo {author} {\bibfnamefont
  {P.}~\bibnamefont {Bertet}}, \bibinfo {author} {\bibfnamefont
  {P.}~\bibnamefont {Roche}}, \ and\ \bibinfo {author} {\bibfnamefont
  {D.}~\bibnamefont {Esteve}},\ }\bibfield  {title} {\enquote {\bibinfo {title}
  {Bright side of the coulomb blockade},}\ }\href {\doibase
  10.1103/PhysRevLett.106.217005} {\bibfield  {journal} {\bibinfo  {journal}
  {Phys. Rev. Lett.}\ }\textbf {\bibinfo {volume} {106}},\ \bibinfo {pages}
  {217005} (\bibinfo {year} {2011})}\BibitemShut {NoStop}%
\bibitem [{\citenamefont {Cottet}\ \emph {et~al.}(2015)\citenamefont {Cottet},
  \citenamefont {Kontos},\ and\ \citenamefont {Dou\ifmmode~\mbox{\c{c}}\else
  \c{c}\fi{}ot}}]{PhysRevB.91.205417}%
  \BibitemOpen
  \bibfield  {author} {\bibinfo {author} {\bibfnamefont {A.}~\bibnamefont
  {Cottet}}, \bibinfo {author} {\bibfnamefont {T.}~\bibnamefont {Kontos}}, \
  and\ \bibinfo {author} {\bibfnamefont {B.}~\bibnamefont
  {Dou\ifmmode~\mbox{\c{c}}\else \c{c}\fi{}ot}},\ }\bibfield  {title} {\enquote
  {\bibinfo {title} {Electron-photon coupling in mesoscopic quantum
  electrodynamics},}\ }\href {\doibase 10.1103/PhysRevB.91.205417} {\bibfield
  {journal} {\bibinfo  {journal} {Phys. Rev. B}\ }\textbf {\bibinfo {volume}
  {91}},\ \bibinfo {pages} {205417} (\bibinfo {year} {2015})}\BibitemShut
  {NoStop}%
\bibitem [{\citenamefont {Dmytruk}\ \emph {et~al.}(2016)\citenamefont
  {Dmytruk}, \citenamefont {Trif}, \citenamefont {Mora},\ and\ \citenamefont
  {Simon}}]{PhysRevB.93.075425}%
  \BibitemOpen
  \bibfield  {author} {\bibinfo {author} {\bibfnamefont {Olesia}\ \bibnamefont
  {Dmytruk}}, \bibinfo {author} {\bibfnamefont {Mircea}\ \bibnamefont {Trif}},
  \bibinfo {author} {\bibfnamefont {Christophe}\ \bibnamefont {Mora}}, \ and\
  \bibinfo {author} {\bibfnamefont {Pascal}\ \bibnamefont {Simon}},\ }\bibfield
   {title} {\enquote {\bibinfo {title} {Out-of-equilibrium quantum dot coupled
  to a microwave cavity},}\ }\href {\doibase 10.1103/PhysRevB.93.075425}
  {\bibfield  {journal} {\bibinfo  {journal} {Phys. Rev. B}\ }\textbf {\bibinfo
  {volume} {93}},\ \bibinfo {pages} {075425} (\bibinfo {year}
  {2016})}\BibitemShut {NoStop}%
\bibitem [{\citenamefont {Mora}\ \emph {et~al.}(2017)\citenamefont {Mora},
  \citenamefont {Altimiras}, \citenamefont {Joyez},\ and\ \citenamefont
  {Portier}}]{PhysRevB.95.125311}%
  \BibitemOpen
  \bibfield  {author} {\bibinfo {author} {\bibfnamefont {C.}~\bibnamefont
  {Mora}}, \bibinfo {author} {\bibfnamefont {C.}~\bibnamefont {Altimiras}},
  \bibinfo {author} {\bibfnamefont {P.}~\bibnamefont {Joyez}}, \ and\ \bibinfo
  {author} {\bibfnamefont {F.}~\bibnamefont {Portier}},\ }\bibfield  {title}
  {\enquote {\bibinfo {title} {Quantum properties of the radiation emitted by a
  conductor in the coulomb blockade regime},}\ }\href {\doibase
  10.1103/PhysRevB.95.125311} {\bibfield  {journal} {\bibinfo  {journal} {Phys.
  Rev. B}\ }\textbf {\bibinfo {volume} {95}},\ \bibinfo {pages} {125311}
  (\bibinfo {year} {2017})}\BibitemShut {NoStop}%
\bibitem [{\citenamefont {Altimiras}\ \emph {et~al.}(2016)\citenamefont
  {Altimiras}, \citenamefont {Portier},\ and\ \citenamefont
  {Joyez}}]{PhysRevX.6.031002}%
  \BibitemOpen
  \bibfield  {author} {\bibinfo {author} {\bibfnamefont {C.}~\bibnamefont
  {Altimiras}}, \bibinfo {author} {\bibfnamefont {F.}~\bibnamefont {Portier}},
  \ and\ \bibinfo {author} {\bibfnamefont {P.}~\bibnamefont {Joyez}},\
  }\bibfield  {title} {\enquote {\bibinfo {title} {Interacting electrodynamics
  of short coherent conductors in quantum circuits},}\ }\href {\doibase
  10.1103/PhysRevX.6.031002} {\bibfield  {journal} {\bibinfo  {journal} {Phys.
  Rev. X}\ }\textbf {\bibinfo {volume} {6}},\ \bibinfo {pages} {031002}
  (\bibinfo {year} {2016})}\BibitemShut {NoStop}%
\bibitem [{\citenamefont {Lepp\"akangas}\ \emph {et~al.}(2014)\citenamefont
  {Lepp\"akangas}, \citenamefont {Johansson}, \citenamefont {Marthaler},\ and\
  \citenamefont {Fogelstr\"om}}]{JuhaNJP}%
  \BibitemOpen
  \bibfield  {author} {\bibinfo {author} {\bibfnamefont {J}~\bibnamefont
  {Lepp\"akangas}}, \bibinfo {author} {\bibfnamefont {G}~\bibnamefont
  {Johansson}}, \bibinfo {author} {\bibfnamefont {M}~\bibnamefont {Marthaler}},
  \ and\ \bibinfo {author} {\bibfnamefont {M}~\bibnamefont {Fogelstr\"om}},\
  }\bibfield  {title} {\enquote {\bibinfo {title} {Input-output description of
  microwave radiation in the dynamical coulomb blockade},}\ }\href
  {http://stacks.iop.org/1367-2630/16/i=1/a=015015} {\bibfield  {journal}
  {\bibinfo  {journal} {New Journal of Physics}\ }\textbf {\bibinfo {volume}
  {16}},\ \bibinfo {pages} {015015} (\bibinfo {year} {2014})}\BibitemShut
  {NoStop}%
\bibitem [{\citenamefont {Lepp\"akangas}\ \emph {et~al.}(2013)\citenamefont
  {Lepp\"akangas}, \citenamefont {Johansson}, \citenamefont {Marthaler},\ and\
  \citenamefont {Fogelstr\"om}}]{PhysRevLett.110.267004}%
  \BibitemOpen
  \bibfield  {author} {\bibinfo {author} {\bibfnamefont {Juha}\ \bibnamefont
  {Lepp\"akangas}}, \bibinfo {author} {\bibfnamefont {G\"oran}\ \bibnamefont
  {Johansson}}, \bibinfo {author} {\bibfnamefont {Michael}\ \bibnamefont
  {Marthaler}}, \ and\ \bibinfo {author} {\bibfnamefont {Mikael}\ \bibnamefont
  {Fogelstr\"om}},\ }\bibfield  {title} {\enquote {\bibinfo {title}
  {Nonclassical photon pair production in a voltage-biased josephson
  junction},}\ }\href {\doibase 10.1103/PhysRevLett.110.267004} {\bibfield
  {journal} {\bibinfo  {journal} {Phys. Rev. Lett.}\ }\textbf {\bibinfo
  {volume} {110}},\ \bibinfo {pages} {267004} (\bibinfo {year}
  {2013})}\BibitemShut {NoStop}%
\bibitem [{\citenamefont {Armour}\ \emph {et~al.}(2013)\citenamefont {Armour},
  \citenamefont {Blencowe}, \citenamefont {Brahimi},\ and\ \citenamefont
  {Rimberg}}]{armour2013universal}%
  \BibitemOpen
  \bibfield  {author} {\bibinfo {author} {\bibfnamefont {A.D.}\ \bibnamefont
  {Armour}}, \bibinfo {author} {\bibfnamefont {M.P.}\ \bibnamefont {Blencowe}},
  \bibinfo {author} {\bibfnamefont {E.}~\bibnamefont {Brahimi}}, \ and\
  \bibinfo {author} {\bibfnamefont {A.J.}\ \bibnamefont {Rimberg}},\ }\bibfield
   {title} {\enquote {\bibinfo {title} {Universal quantum fluctuations of a
  cavity mode driven by a josephson junction},}\ }\href@noop {} {\bibfield
  {journal} {\bibinfo  {journal} {Physical review letters}\ }\textbf {\bibinfo
  {volume} {111}},\ \bibinfo {pages} {247001} (\bibinfo {year}
  {2013})}\BibitemShut {NoStop}%
\bibitem [{\citenamefont {Gramich}\ \emph {et~al.}(2013)\citenamefont
  {Gramich}, \citenamefont {Kubala}, \citenamefont {Rohrer},\ and\
  \citenamefont {Ankerhold}}]{PhysRevLett.111.247002}%
  \BibitemOpen
  \bibfield  {author} {\bibinfo {author} {\bibfnamefont {Vera}\ \bibnamefont
  {Gramich}}, \bibinfo {author} {\bibfnamefont {Bj\"orn}\ \bibnamefont
  {Kubala}}, \bibinfo {author} {\bibfnamefont {Selina}\ \bibnamefont {Rohrer}},
  \ and\ \bibinfo {author} {\bibfnamefont {Joachim}\ \bibnamefont
  {Ankerhold}},\ }\bibfield  {title} {\enquote {\bibinfo {title} {From
  coulomb-blockade to nonlinear quantum dynamics in a superconducting circuit
  with a resonator},}\ }\href {\doibase 10.1103/PhysRevLett.111.247002}
  {\bibfield  {journal} {\bibinfo  {journal} {Phys. Rev. Lett.}\ }\textbf
  {\bibinfo {volume} {111}},\ \bibinfo {pages} {247002} (\bibinfo {year}
  {2013})}\BibitemShut {NoStop}%
\bibitem [{\citenamefont {Beenakker}\ and\ \citenamefont
  {Schomerus}(2001)}]{PhysRevLett.86.700}%
  \BibitemOpen
  \bibfield  {author} {\bibinfo {author} {\bibfnamefont {C.~W.~J.}\
  \bibnamefont {Beenakker}}\ and\ \bibinfo {author} {\bibfnamefont
  {H.}~\bibnamefont {Schomerus}},\ }\bibfield  {title} {\enquote {\bibinfo
  {title} {Counting statistics of photons produced by electronic shot noise},}\
  }\href {\doibase 10.1103/PhysRevLett.86.700} {\bibfield  {journal} {\bibinfo
  {journal} {Phys. Rev. Lett.}\ }\textbf {\bibinfo {volume} {86}},\ \bibinfo
  {pages} {700--703} (\bibinfo {year} {2001})}\BibitemShut {NoStop}%
\bibitem [{\citenamefont {Beenakker}\ and\ \citenamefont
  {Schomerus}(2004)}]{PhysRevLett.93.096801}%
  \BibitemOpen
  \bibfield  {author} {\bibinfo {author} {\bibfnamefont {C.~W.~J.}\
  \bibnamefont {Beenakker}}\ and\ \bibinfo {author} {\bibfnamefont
  {H.}~\bibnamefont {Schomerus}},\ }\bibfield  {title} {\enquote {\bibinfo
  {title} {Antibunched photons emitted by a quantum point contact out of
  equilibrium},}\ }\href {\doibase 10.1103/PhysRevLett.93.096801} {\bibfield
  {journal} {\bibinfo  {journal} {Phys. Rev. Lett.}\ }\textbf {\bibinfo
  {volume} {93}},\ \bibinfo {pages} {096801} (\bibinfo {year}
  {2004})}\BibitemShut {NoStop}%
\bibitem [{\citenamefont {Lebedev}\ \emph {et~al.}(2010)\citenamefont
  {Lebedev}, \citenamefont {Lesovik},\ and\ \citenamefont
  {Blatter}}]{PhysRevB.81.155421}%
  \BibitemOpen
  \bibfield  {author} {\bibinfo {author} {\bibfnamefont {A.~V.}\ \bibnamefont
  {Lebedev}}, \bibinfo {author} {\bibfnamefont {G.~B.}\ \bibnamefont
  {Lesovik}}, \ and\ \bibinfo {author} {\bibfnamefont {G.}~\bibnamefont
  {Blatter}},\ }\bibfield  {title} {\enquote {\bibinfo {title} {Statistics of
  radiation emitted from a quantum point contact},}\ }\href {\doibase
  10.1103/PhysRevB.81.155421} {\bibfield  {journal} {\bibinfo  {journal} {Phys.
  Rev. B}\ }\textbf {\bibinfo {volume} {81}},\ \bibinfo {pages} {155421}
  (\bibinfo {year} {2010})}\BibitemShut {NoStop}%
\bibitem [{\citenamefont {Fulga}\ \emph {et~al.}(2010)\citenamefont {Fulga},
  \citenamefont {Hassler},\ and\ \citenamefont
  {Beenakker}}]{PhysRevB.81.115331}%
  \BibitemOpen
  \bibfield  {author} {\bibinfo {author} {\bibfnamefont {I.~C.}\ \bibnamefont
  {Fulga}}, \bibinfo {author} {\bibfnamefont {F.}~\bibnamefont {Hassler}}, \
  and\ \bibinfo {author} {\bibfnamefont {C.~W.~J.}\ \bibnamefont {Beenakker}},\
  }\bibfield  {title} {\enquote {\bibinfo {title} {Nonzero temperature effects
  on antibunched photons emitted by a quantum point contact out of
  equilibrium},}\ }\href {\doibase 10.1103/PhysRevB.81.115331} {\bibfield
  {journal} {\bibinfo  {journal} {Phys. Rev. B}\ }\textbf {\bibinfo {volume}
  {81}},\ \bibinfo {pages} {115331} (\bibinfo {year} {2010})}\BibitemShut
  {NoStop}%
\bibitem [{\citenamefont {Hassler}\ and\ \citenamefont
  {Otten}(2015)}]{PhysRevB.92.195417}%
  \BibitemOpen
  \bibfield  {author} {\bibinfo {author} {\bibfnamefont {Fabian}\ \bibnamefont
  {Hassler}}\ and\ \bibinfo {author} {\bibfnamefont {Daniel}\ \bibnamefont
  {Otten}},\ }\bibfield  {title} {\enquote {\bibinfo {title} {Second-order
  coherence of microwave photons emitted by a quantum point contact},}\ }\href
  {\doibase 10.1103/PhysRevB.92.195417} {\bibfield  {journal} {\bibinfo
  {journal} {Phys. Rev. B}\ }\textbf {\bibinfo {volume} {92}},\ \bibinfo
  {pages} {195417} (\bibinfo {year} {2015})}\BibitemShut {NoStop}%
\bibitem [{\citenamefont {Grimm}\ \emph {et~al.}(2019)\citenamefont {Grimm},
  \citenamefont {Blanchet}, \citenamefont {Albert}, \citenamefont
  {Lepp\"akangas}, \citenamefont {Jebari}, \citenamefont {Hazra}, \citenamefont
  {Gustavo}, \citenamefont {Thomassin}, \citenamefont {Dupont-Ferrier},
  \citenamefont {Portier},\ and\ \citenamefont {Hofheinz}}]{PhysRevX.9.021016}%
  \BibitemOpen
  \bibfield  {author} {\bibinfo {author} {\bibfnamefont {A.}~\bibnamefont
  {Grimm}}, \bibinfo {author} {\bibfnamefont {F.}~\bibnamefont {Blanchet}},
  \bibinfo {author} {\bibfnamefont {R.}~\bibnamefont {Albert}}, \bibinfo
  {author} {\bibfnamefont {J.}~\bibnamefont {Lepp\"akangas}}, \bibinfo {author}
  {\bibfnamefont {S.}~\bibnamefont {Jebari}}, \bibinfo {author} {\bibfnamefont
  {D.}~\bibnamefont {Hazra}}, \bibinfo {author} {\bibfnamefont
  {F.}~\bibnamefont {Gustavo}}, \bibinfo {author} {\bibfnamefont {J.-L.}\
  \bibnamefont {Thomassin}}, \bibinfo {author} {\bibfnamefont {E.}~\bibnamefont
  {Dupont-Ferrier}}, \bibinfo {author} {\bibfnamefont {F.}~\bibnamefont
  {Portier}}, \ and\ \bibinfo {author} {\bibfnamefont {M.}~\bibnamefont
  {Hofheinz}},\ }\bibfield  {title} {\enquote {\bibinfo {title} {Bright
  on-demand source of antibunched microwave photons based on inelastic cooper
  pair tunneling},}\ }\href {\doibase 10.1103/PhysRevX.9.021016} {\bibfield
  {journal} {\bibinfo  {journal} {Phys. Rev. X}\ }\textbf {\bibinfo {volume}
  {9}},\ \bibinfo {pages} {021016} (\bibinfo {year} {2019})}\BibitemShut
  {NoStop}%
\bibitem [{\citenamefont {Rolland}\ \emph {et~al.}(2019)\citenamefont
  {Rolland}, \citenamefont {Peugeot}, \citenamefont {Dambach}, \citenamefont
  {Westig}, \citenamefont {Kubala}, \citenamefont {Mukharsky}, \citenamefont
  {Altimiras}, \citenamefont {le~Sueur}, \citenamefont {Joyez}, \citenamefont
  {Vion}, \citenamefont {Roche}, \citenamefont {Esteve}, \citenamefont
  {Ankerhold},\ and\ \citenamefont {Portier}}]{PhysRevLett.122.186804}%
  \BibitemOpen
  \bibfield  {author} {\bibinfo {author} {\bibfnamefont {C.}~\bibnamefont
  {Rolland}}, \bibinfo {author} {\bibfnamefont {A.}~\bibnamefont {Peugeot}},
  \bibinfo {author} {\bibfnamefont {S.}~\bibnamefont {Dambach}}, \bibinfo
  {author} {\bibfnamefont {M.}~\bibnamefont {Westig}}, \bibinfo {author}
  {\bibfnamefont {B.}~\bibnamefont {Kubala}}, \bibinfo {author} {\bibfnamefont
  {Y.}~\bibnamefont {Mukharsky}}, \bibinfo {author} {\bibfnamefont
  {C.}~\bibnamefont {Altimiras}}, \bibinfo {author} {\bibfnamefont
  {H.}~\bibnamefont {le~Sueur}}, \bibinfo {author} {\bibfnamefont
  {P.}~\bibnamefont {Joyez}}, \bibinfo {author} {\bibfnamefont
  {D.}~\bibnamefont {Vion}}, \bibinfo {author} {\bibfnamefont {P.}~\bibnamefont
  {Roche}}, \bibinfo {author} {\bibfnamefont {D.}~\bibnamefont {Esteve}},
  \bibinfo {author} {\bibfnamefont {J.}~\bibnamefont {Ankerhold}}, \ and\
  \bibinfo {author} {\bibfnamefont {F.}~\bibnamefont {Portier}},\ }\bibfield
  {title} {\enquote {\bibinfo {title} {Antibunched photons emitted by a
  dc-biased josephson junction},}\ }\href {\doibase
  10.1103/PhysRevLett.122.186804} {\bibfield  {journal} {\bibinfo  {journal}
  {Phys. Rev. Lett.}\ }\textbf {\bibinfo {volume} {122}},\ \bibinfo {pages}
  {186804} (\bibinfo {year} {2019})}\BibitemShut {NoStop}%
\bibitem [{\citenamefont {Cassidy}\ \emph {et~al.}(2017)\citenamefont
  {Cassidy}, \citenamefont {Bruno}, \citenamefont {Rubbert}, \citenamefont
  {Irfan}, \citenamefont {Kammhuber}, \citenamefont {Schouten}, \citenamefont
  {Akhmerov},\ and\ \citenamefont {Kouwenhoven}}]{cassidy2017}%
  \BibitemOpen
  \bibfield  {author} {\bibinfo {author} {\bibfnamefont {M.~C.}\ \bibnamefont
  {Cassidy}}, \bibinfo {author} {\bibfnamefont {A.}~\bibnamefont {Bruno}},
  \bibinfo {author} {\bibfnamefont {S.}~\bibnamefont {Rubbert}}, \bibinfo
  {author} {\bibfnamefont {M.}~\bibnamefont {Irfan}}, \bibinfo {author}
  {\bibfnamefont {J.}~\bibnamefont {Kammhuber}}, \bibinfo {author}
  {\bibfnamefont {R.~N.}\ \bibnamefont {Schouten}}, \bibinfo {author}
  {\bibfnamefont {A.~R.}\ \bibnamefont {Akhmerov}}, \ and\ \bibinfo {author}
  {\bibfnamefont {L.~P.}\ \bibnamefont {Kouwenhoven}},\ }\bibfield  {title}
  {\enquote {\bibinfo {title} {Demonstration of an ac josephson junction
  laser},}\ }\href {\doibase 10.1126/science.aah6640} {\bibfield  {journal}
  {\bibinfo  {journal} {Science}\ }\textbf {\bibinfo {volume} {355}},\ \bibinfo
  {pages} {939--942} (\bibinfo {year} {2017})}\BibitemShut {NoStop}%
\bibitem [{\citenamefont {Godschalk}\ \emph {et~al.}(2011)\citenamefont
  {Godschalk}, \citenamefont {Hassler},\ and\ \citenamefont
  {Nazarov}}]{PhysRevLett.107.073901}%
  \BibitemOpen
  \bibfield  {author} {\bibinfo {author} {\bibfnamefont {Frans}\ \bibnamefont
  {Godschalk}}, \bibinfo {author} {\bibfnamefont {Fabian}\ \bibnamefont
  {Hassler}}, \ and\ \bibinfo {author} {\bibfnamefont {Yuli~V.}\ \bibnamefont
  {Nazarov}},\ }\bibfield  {title} {\enquote {\bibinfo {title} {Proposal for an
  optical laser producing light at half the josephson frequency},}\ }\href
  {\doibase 10.1103/PhysRevLett.107.073901} {\bibfield  {journal} {\bibinfo
  {journal} {Phys. Rev. Lett.}\ }\textbf {\bibinfo {volume} {107}},\ \bibinfo
  {pages} {073901} (\bibinfo {year} {2011})}\BibitemShut {NoStop}%
\bibitem [{\citenamefont {Godschalk}\ and\ \citenamefont
  {Nazarov}(2013)}]{PhysRevB.87.094511}%
  \BibitemOpen
  \bibfield  {author} {\bibinfo {author} {\bibfnamefont {Frans}\ \bibnamefont
  {Godschalk}}\ and\ \bibinfo {author} {\bibfnamefont {Yuli~V.}\ \bibnamefont
  {Nazarov}},\ }\bibfield  {title} {\enquote {\bibinfo {title} {Lasing at half
  the josephson frequency with exponentially long coherence times},}\ }\href
  {\doibase 10.1103/PhysRevB.87.094511} {\bibfield  {journal} {\bibinfo
  {journal} {Phys. Rev. B}\ }\textbf {\bibinfo {volume} {87}},\ \bibinfo
  {pages} {094511} (\bibinfo {year} {2013})}\BibitemShut {NoStop}%
\bibitem [{\citenamefont {Godschalk}\ and\ \citenamefont
  {Nazarov}(2014)}]{PhysRevB.89.104502}%
  \BibitemOpen
  \bibfield  {author} {\bibinfo {author} {\bibfnamefont {Frans}\ \bibnamefont
  {Godschalk}}\ and\ \bibinfo {author} {\bibfnamefont {Yuli~V.}\ \bibnamefont
  {Nazarov}},\ }\bibfield  {title} {\enquote {\bibinfo {title}
  {Light-superconducting interference devices},}\ }\href {\doibase
  10.1103/PhysRevB.89.104502} {\bibfield  {journal} {\bibinfo  {journal} {Phys.
  Rev. B}\ }\textbf {\bibinfo {volume} {89}},\ \bibinfo {pages} {104502}
  (\bibinfo {year} {2014})}\BibitemShut {NoStop}%
\bibitem [{\citenamefont {Jebari}\ \emph {et~al.}(2018)\citenamefont {Jebari},
  \citenamefont {Blanchet}, \citenamefont {Grimm}, \citenamefont {Hazra},
  \citenamefont {Albert}, \citenamefont {Joyez}, \citenamefont {Vion},
  \citenamefont {Estève}, \citenamefont {Portier},\ and\ \citenamefont
  {Hofheinz}}]{Jebari18}%
  \BibitemOpen
  \bibfield  {author} {\bibinfo {author} {\bibfnamefont {S.}~\bibnamefont
  {Jebari}}, \bibinfo {author} {\bibfnamefont {F.}~\bibnamefont {Blanchet}},
  \bibinfo {author} {\bibfnamefont {A.}~\bibnamefont {Grimm}}, \bibinfo
  {author} {\bibfnamefont {D.}~\bibnamefont {Hazra}}, \bibinfo {author}
  {\bibfnamefont {R.}~\bibnamefont {Albert}}, \bibinfo {author} {\bibfnamefont
  {P.}~\bibnamefont {Joyez}}, \bibinfo {author} {\bibfnamefont
  {D.}~\bibnamefont {Vion}}, \bibinfo {author} {\bibfnamefont {D.}~\bibnamefont
  {Estève}}, \bibinfo {author} {\bibfnamefont {F.}~\bibnamefont {Portier}}, \
  and\ \bibinfo {author} {\bibfnamefont {M.}~\bibnamefont {Hofheinz}},\
  }\bibfield  {title} {\enquote {\bibinfo {title} {Near-quantum-limited
  amplification from inelastic cooper-pair tunnelling},}\ }\href {\doibase
  10.1038/s41928-018-0055-7} {\bibfield  {journal} {\bibinfo  {journal} {Nature
  Electronics}\ }\textbf {\bibinfo {volume} {1}},\ \bibinfo {pages} {223--227}
  (\bibinfo {year} {2018})}\BibitemShut {NoStop}%
\bibitem [{\citenamefont {Mendes}\ \emph {et~al.}(2019)\citenamefont {Mendes},
  \citenamefont {Jezouin}, \citenamefont {Joyez}, \citenamefont {Reulet},
  \citenamefont {Blais}, \citenamefont {Portier}, \citenamefont {Mora},\ and\
  \citenamefont {Altimiras}}]{PhysRevApplied.11.034035}%
  \BibitemOpen
  \bibfield  {author} {\bibinfo {author} {\bibfnamefont {Udson~C.}\
  \bibnamefont {Mendes}}, \bibinfo {author} {\bibfnamefont {S\'ebastien}\
  \bibnamefont {Jezouin}}, \bibinfo {author} {\bibfnamefont {Philippe}\
  \bibnamefont {Joyez}}, \bibinfo {author} {\bibfnamefont {Bertrand}\
  \bibnamefont {Reulet}}, \bibinfo {author} {\bibfnamefont {Alexandre}\
  \bibnamefont {Blais}}, \bibinfo {author} {\bibfnamefont {Fabien}\
  \bibnamefont {Portier}}, \bibinfo {author} {\bibfnamefont {Christophe}\
  \bibnamefont {Mora}}, \ and\ \bibinfo {author} {\bibfnamefont {Carles}\
  \bibnamefont {Altimiras}},\ }\bibfield  {title} {\enquote {\bibinfo {title}
  {Parametric amplification and squeezing with an ac- and dc-voltage biased
  superconducting junction},}\ }\href {\doibase
  10.1103/PhysRevApplied.11.034035} {\bibfield  {journal} {\bibinfo  {journal}
  {Phys. Rev. Applied}\ }\textbf {\bibinfo {volume} {11}},\ \bibinfo {pages}
  {034035} (\bibinfo {year} {2019})}\BibitemShut {NoStop}%
\bibitem [{\citenamefont {Forgues}\ \emph {et~al.}(2015)\citenamefont
  {Forgues}, \citenamefont {Lupien},\ and\ \citenamefont
  {Reulet}}]{PhysRevLett.114.130403}%
  \BibitemOpen
  \bibfield  {author} {\bibinfo {author} {\bibfnamefont {Jean-Charles}\
  \bibnamefont {Forgues}}, \bibinfo {author} {\bibfnamefont {Christian}\
  \bibnamefont {Lupien}}, \ and\ \bibinfo {author} {\bibfnamefont {Bertrand}\
  \bibnamefont {Reulet}},\ }\bibfield  {title} {\enquote {\bibinfo {title}
  {Experimental violation of bell-like inequalities by electronic shot
  noise},}\ }\href {\doibase 10.1103/PhysRevLett.114.130403} {\bibfield
  {journal} {\bibinfo  {journal} {Phys. Rev. Lett.}\ }\textbf {\bibinfo
  {volume} {114}},\ \bibinfo {pages} {130403} (\bibinfo {year}
  {2015})}\BibitemShut {NoStop}%
\bibitem [{\citenamefont {Cottet}\ \emph {et~al.}(2020)\citenamefont {Cottet},
  \citenamefont {Leghtas},\ and\ \citenamefont {Kontos}}]{PhysRevB.102.155105}%
  \BibitemOpen
  \bibfield  {author} {\bibinfo {author} {\bibfnamefont {Audrey}\ \bibnamefont
  {Cottet}}, \bibinfo {author} {\bibfnamefont {Zaki}\ \bibnamefont {Leghtas}},
  \ and\ \bibinfo {author} {\bibfnamefont {Takis}\ \bibnamefont {Kontos}},\
  }\bibfield  {title} {\enquote {\bibinfo {title} {Theory of interactions
  between cavity photons induced by a mesoscopic circuit},}\ }\href {\doibase
  10.1103/PhysRevB.102.155105} {\bibfield  {journal} {\bibinfo  {journal}
  {Phys. Rev. B}\ }\textbf {\bibinfo {volume} {102}},\ \bibinfo {pages}
  {155105} (\bibinfo {year} {2020})}\BibitemShut {NoStop}%
\bibitem [{\citenamefont {Peugeot}\ \emph {et~al.}(2021)\citenamefont
  {Peugeot}, \citenamefont {Ménard}, \citenamefont {Dambach}, \citenamefont
  {Westig}, \citenamefont {Kubala}, \citenamefont {Mukharsky}, \citenamefont
  {Altimiras}, \citenamefont {Joyez}, \citenamefont {Vion}, \citenamefont
  {Roche}, \citenamefont {Esteve}, \citenamefont {Milman}, \citenamefont
  {Lepp\"akangas}, \citenamefont {Johansson}, \citenamefont {Hofheinz},
  \citenamefont {Ankerhold},\ and\ \citenamefont {Portier}}]{Peugeot2021}%
  \BibitemOpen
  \bibfield  {author} {\bibinfo {author} {\bibfnamefont {A.}~\bibnamefont
  {Peugeot}}, \bibinfo {author} {\bibfnamefont {G.}~\bibnamefont {Ménard}},
  \bibinfo {author} {\bibfnamefont {S.}~\bibnamefont {Dambach}}, \bibinfo
  {author} {\bibfnamefont {M.}~\bibnamefont {Westig}}, \bibinfo {author}
  {\bibfnamefont {B.}~\bibnamefont {Kubala}}, \bibinfo {author} {\bibfnamefont
  {Y.}~\bibnamefont {Mukharsky}}, \bibinfo {author} {\bibfnamefont
  {C.}~\bibnamefont {Altimiras}}, \bibinfo {author} {\bibfnamefont
  {P.}~\bibnamefont {Joyez}}, \bibinfo {author} {\bibfnamefont
  {D.}~\bibnamefont {Vion}}, \bibinfo {author} {\bibfnamefont {P.}~\bibnamefont
  {Roche}}, \bibinfo {author} {\bibfnamefont {D.}~\bibnamefont {Esteve}},
  \bibinfo {author} {\bibfnamefont {P.}~\bibnamefont {Milman}}, \bibinfo
  {author} {\bibfnamefont {J.}~\bibnamefont {Lepp\"akangas}}, \bibinfo {author}
  {\bibfnamefont {G.}~\bibnamefont {Johansson}}, \bibinfo {author}
  {\bibfnamefont {M.}~\bibnamefont {Hofheinz}}, \bibinfo {author}
  {\bibfnamefont {J.}~\bibnamefont {Ankerhold}}, \ and\ \bibinfo {author}
  {\bibfnamefont {F.}~\bibnamefont {Portier}},\ }\bibfield  {title} {\enquote
  {\bibinfo {title} {Generating two continuous entangled microwave beams using
  a dc-biased josephson junction},}\ }\href@noop {} {\bibfield  {journal}
  {\bibinfo  {journal} {Phys. Rev. X}\ }\textbf {\bibinfo {volume} {11}},\
  \bibinfo {pages} {031008} (\bibinfo {year} {2021})}\BibitemShut {NoStop}%
\bibitem [{\citenamefont {Est\`eve}\ \emph {et~al.}(2018)\citenamefont
  {Est\`eve}, \citenamefont {Aprili},\ and\ \citenamefont
  {Gabelli}}]{1807.02364}%
  \BibitemOpen
  \bibfield  {author} {\bibinfo {author} {\bibfnamefont {J\'er\^ome}\
  \bibnamefont {Est\`eve}}, \bibinfo {author} {\bibfnamefont {Marco}\
  \bibnamefont {Aprili}}, \ and\ \bibinfo {author} {\bibfnamefont {Julien}\
  \bibnamefont {Gabelli}},\ }\href@noop {} {\enquote {\bibinfo {title} {Quantum
  dynamics of a microwave resonator strongly coupled to a tunnel junction},}\ }
  (\bibinfo {year} {2018}),\ \Eprint {http://arxiv.org/abs/arXiv:1807.02364}
  {arXiv:1807.02364} \BibitemShut {NoStop}%
\bibitem [{\citenamefont {Aiello}(2021)}]{AielloPhD}%
  \BibitemOpen
  \bibfield  {author} {\bibinfo {author} {\bibfnamefont {Gianluca}\
  \bibnamefont {Aiello}},\ }\emph {\bibinfo {title} {Quantum dynamics of a high
  impedance cavity strongly coupled to a Josephson junction}},\ \href@noop {}
  {Ph.D. thesis},\ \bibinfo  {school} {Paris-Saclay University} (\bibinfo
  {year} {2021})\BibitemShut {NoStop}%
\bibitem [{\citenamefont {Dambach}\ \emph {et~al.}(2015)\citenamefont
  {Dambach}, \citenamefont {Kubala}, \citenamefont {Gramich},\ and\
  \citenamefont {Ankerhold}}]{PhysRevB.92.054508}%
  \BibitemOpen
  \bibfield  {author} {\bibinfo {author} {\bibfnamefont {Simon}\ \bibnamefont
  {Dambach}}, \bibinfo {author} {\bibfnamefont {Bjorn}\ \bibnamefont {Kubala}},
  \bibinfo {author} {\bibfnamefont {Vera}\ \bibnamefont {Gramich}}, \ and\
  \bibinfo {author} {\bibfnamefont {Joachim}\ \bibnamefont {Ankerhold}},\
  }\bibfield  {title} {\enquote {\bibinfo {title} {Time-resolved statistics of
  nonclassical light in josephson photonics},}\ }\href {\doibase
  10.1103/PhysRevB.92.054508} {\bibfield  {journal} {\bibinfo  {journal} {Phys.
  Rev. B}\ }\textbf {\bibinfo {volume} {92}},\ \bibinfo {pages} {054508}
  (\bibinfo {year} {2015})}\BibitemShut {NoStop}%
\bibitem [{\citenamefont {Hofer}\ \emph {et~al.}(2016)\citenamefont {Hofer},
  \citenamefont {Souquet},\ and\ \citenamefont
  {Clerk}}]{PhysRevB.93.041418(R)}%
  \BibitemOpen
  \bibfield  {author} {\bibinfo {author} {\bibfnamefont {Patrick~P.}\
  \bibnamefont {Hofer}}, \bibinfo {author} {\bibfnamefont {J.-R.}\ \bibnamefont
  {Souquet}}, \ and\ \bibinfo {author} {\bibfnamefont {A.~A.}\ \bibnamefont
  {Clerk}},\ }\bibfield  {title} {\enquote {\bibinfo {title} {Quantum heat
  engine based on photon-assisted cooper pair tunneling},}\ }\href {\doibase
  doi.org/10.1103/PhysRevB.93.041418} {\bibfield  {journal} {\bibinfo
  {journal} {Phys. Rev. B}\ }\textbf {\bibinfo {volume} {93}},\ \bibinfo
  {pages} {041418} (\bibinfo {year} {2016})}\BibitemShut {NoStop}%
\bibitem [{\citenamefont {Emary}\ \emph {et~al.}(2012)\citenamefont {Emary},
  \citenamefont {Pöltl}, \citenamefont {Carmele}, \citenamefont {Kabuss},
  \citenamefont {Knorr},\ and\ \citenamefont {Brandes}}]{g2ToFano}%
  \BibitemOpen
  \bibfield  {author} {\bibinfo {author} {\bibfnamefont {C.}~\bibnamefont
  {Emary}}, \bibinfo {author} {\bibfnamefont {C.}~\bibnamefont {Pöltl}},
  \bibinfo {author} {\bibfnamefont {A.}~\bibnamefont {Carmele}}, \bibinfo
  {author} {\bibfnamefont {J.}~\bibnamefont {Kabuss}}, \bibinfo {author}
  {\bibfnamefont {A.}~\bibnamefont {Knorr}}, \ and\ \bibinfo {author}
  {\bibfnamefont {T.}~\bibnamefont {Brandes}},\ }\bibfield  {title} {\enquote
  {\bibinfo {title} {Bunching and antibunching in electronic transport},}\
  }\href@noop {} {\bibfield  {journal} {\bibinfo  {journal} {Physical Review
  B}\ }\textbf {\bibinfo {volume} {85}},\ \bibinfo {pages} {165417} (\bibinfo
  {year} {2012})}\BibitemShut {NoStop}%
\bibitem [{\citenamefont {Armour}\ \emph {et~al.}(2017)\citenamefont {Armour},
  \citenamefont {Kubala},\ and\ \citenamefont {Ankerhold}}]{Armour2017}%
  \BibitemOpen
  \bibfield  {author} {\bibinfo {author} {\bibfnamefont {Andrew~D.}\
  \bibnamefont {Armour}}, \bibinfo {author} {\bibfnamefont {Bj\"orn}\
  \bibnamefont {Kubala}}, \ and\ \bibinfo {author} {\bibfnamefont {Joachim}\
  \bibnamefont {Ankerhold}},\ }\bibfield  {title} {\enquote {\bibinfo {title}
  {Noise switching at a dynamical critical point in a cavity-conductor
  hybrid},}\ }\href {\doibase 10.1103/PhysRevB.96.214509} {\bibfield  {journal}
  {\bibinfo  {journal} {Phys. Rev. B}\ }\textbf {\bibinfo {volume} {96}},\
  \bibinfo {pages} {214509} (\bibinfo {year} {2017})}\BibitemShut {NoStop}%
\bibitem [{\citenamefont {N}(2009)}]{HofheinzMartinis}%
  \BibitemOpen
  \bibfield  {author} {\bibinfo {author} {\bibfnamefont {Hofheinz M Wang H
  Ansmann M Bialczak RC Lucero E Neeley M O'Connell AD Sank D Wenner J Martinis
  JM Cleland~A}\ \bibnamefont {N}},\ }\bibfield  {title} {\enquote {\bibinfo
  {title} {Synthesizing arbitrary quantum states in a superconducting
  resonator},}\ }\href@noop {} {\bibfield  {journal} {\bibinfo  {journal}
  {Nature}\ }\textbf {\bibinfo {volume} {459}},\ \bibinfo {pages} {546}
  (\bibinfo {year} {2009})}\BibitemShut {NoStop}%
\bibitem [{\citenamefont {L.}(2015)}]{Krastanov}%
  \BibitemOpen
  \bibfield  {author} {\bibinfo {author} {\bibfnamefont {Krastanov S. Albert V.
  V. Shen C. Zou CL. Heeres R.W. Vlastakis B. Schoelkopf R.J.~Jiang}\
  \bibnamefont {L.}},\ }\bibfield  {title} {\enquote {\bibinfo {title}
  {Universal control of an oscillator with dispersive coupling to a qubit},}\
  }\href@noop {} {\bibfield  {journal} {\bibinfo  {journal} {Phys. Rev. A}\
  }\textbf {\bibinfo {volume} {92}},\ \bibinfo {pages} {040303} (\bibinfo
  {year} {2015})}\BibitemShut {NoStop}%
\bibitem [{\citenamefont {Padurariu}\ \emph {et~al.}(2012)\citenamefont
  {Padurariu}, \citenamefont {Hassler},\ and\ \citenamefont
  {Nazarov}}]{PhysRevB.86.054514}%
  \BibitemOpen
  \bibfield  {author} {\bibinfo {author} {\bibfnamefont {Ciprian}\ \bibnamefont
  {Padurariu}}, \bibinfo {author} {\bibfnamefont {Fabian}\ \bibnamefont
  {Hassler}}, \ and\ \bibinfo {author} {\bibfnamefont {Yuli~V.}\ \bibnamefont
  {Nazarov}},\ }\bibfield  {title} {\enquote {\bibinfo {title} {Statistics of
  radiation at josephson parametric resonance},}\ }\href {\doibase
  10.1103/PhysRevB.86.054514} {\bibfield  {journal} {\bibinfo  {journal} {Phys.
  Rev. B}\ }\textbf {\bibinfo {volume} {86}},\ \bibinfo {pages} {054514}
  (\bibinfo {year} {2012})}\BibitemShut {NoStop}%
\bibitem [{\citenamefont {Lang}\ and\ \citenamefont
  {Armour}(2021)}]{LangNJP2021}%
  \BibitemOpen
  \bibfield  {author} {\bibinfo {author} {\bibfnamefont {B.}~\bibnamefont
  {Lang}}\ and\ \bibinfo {author} {\bibfnamefont {A.~D.}\ \bibnamefont
  {Armour}},\ }\href@noop {} {\bibfield  {journal} {\bibinfo  {journal} {New J.
  Phys.}\ }\textbf {\bibinfo {volume} {23}},\ \bibinfo {pages} {033021}
  (\bibinfo {year} {2021})}\BibitemShut {NoStop}%
\end{thebibliography}%

\end{document}